\documentclass[reprint,10pt,twocolumn]{revtex4-1}

 \usepackage{graphicx}
\usepackage[colorlinks=true]{hyperref} 
\usepackage[usenames,dvipsnames]{color}
\usepackage{subcaption}
\usepackage{float}

\usepackage{amsmath,amssymb,amsbsy}
\usepackage{bbm,mathrsfs,yfonts,eufrak}
\renewcommand{\vec}[1]{\boldsymbol{#1}}

\begin{document}
\preprint{}

 \title{The influence of temperature dynamics and dynamic finite ion Larmor radius effects on seeded high amplitude plasma blobs}
\author{M. Held }
\email{markus.held@uibk.ac.at}
\author{M. Wiesenberger }
\affiliation{Institute for Ion Physics and Applied Physics, 
 Universit{\"a}t Innsbruck, A-6020 Innsbruck, Austria}
 \author{J. Madsen }
\affiliation{Department of Physics, Technical University of Denmark, DK-2800 Kgs. Lyngby, Denmark}
\author{A. Kendl}
\affiliation{Institute for Ion Physics and Applied Physics, 
 Universit{\"a}t Innsbruck, A-6020 Innsbruck, Austria}

\begin{abstract}
Thermal effects on the perpendicular convection of seeded pressure blobs in the scrape-off layer of magnetised fusion plasmas 
are investigated. Our numerical study is based on a four field full-F gyrofluid model, which 
entails the consistent description of high fluctuation amplitudes and dynamic finite Larmor radius effects. 
We find that the maximal radial blob velocity increases with the square root of the initial pressure perturbation 
and that a finite Larmor radius contributes to highly compact blob structures that propagate in the poloidal direction.
An extensive parameter study reveals that a smooth transition to this compact blob regime occurs when 
the finite Larmor radius effect strength, defined by the ratio of the magnetic field aligned component of the ion diamagnetic to the 
$\vec{E}\times\vec{B}$ vorticity, exceeds unity.
The maximal radial blob velocities agree excellently with the inertial velocity scaling law over 
more than an order of magnitude. 
We show that the finite Larmor radius effect strength affects the poloidal and total particle transport and 
present an empirical scaling law for the poloidal and total blob velocities.
Distinctions to the blob behaviour in the isothermal limit with constant finite Larmor radius.
\end{abstract}
\maketitle 
\section{Introduction}\label{sec:intro}
In magnetised fusion plasmas the exhaust of particles and heat critically determines the lifetime of the plasma facing components.
Intermittent outbreaks of spatially localised structures~\cite{antar01,antar03} may damage the divertor plates and the surrounding wall~\cite{rudakov05,loarte07,rudakov09}.
These structures, often referred to as blobs or filaments, are born in the edge in the vicinity of the last closed flux surface (LCFS) and are expelled into the scrape-off layer (SOL).
\\
The strong magnetic field separates the spatial scales of filaments to the extent that they are strongly elongated
along the magnetic field but spatially localised perpendicular to it. 
Typically their cross-field size $\sigma$ is on the order of $cm$, the perpendicular ion Mach number is subsonic and the
relative fluctuation levels of particle density and temperature in the SOL are of order unity~\cite{birkenmeier15,endler95,zweben15}. 
In the SOL parallel heat conduction cools electrons much faster than ions and leads to ion to electron background temperature ratios in the range of
$\tau_i\equiv t_{i0}/t_{e0}\sim 1-4$~\cite{kocan12,elmore12}. 
\\
The radial transport caused by blobs depends on their production rate and the blob regime. 
A blob regime is a parameter space, in which a particular physical mechanism dominates the dynamics of the blob. 
The various blob regimes strongly depend on the blobs' cross-field size and collisionality~\cite{dippolito11,manz13}.
In all regimes it is the magnetic field inhomogeneity that generates perpendicular $\vec{E}\times\vec{B}$ energy. 
The mechanism for radial motion is based on the curvature and $\vec{\nabla} B$ drifts, which induce an electrical field in the poloidal direction. The resulting 
$\vec{E}\times\vec{B}$ drift moves the blob radially outward. 
One particular regime is the so called inertial or resistive ballooning regime, predominantly emerging when the filament is disconnected from the divertor plates due to high SOL collisionality~\cite{carralero15,myra06}, 
high magnetic shear in the vicinity of the X-point or electromagnetic effects~\cite{krasheninnikov08}.
In the inertial regime the basic blob dynamics are captured by two dimensional models that invoke the cold ion temperature and thin-layer (Boussinesq) approximation~\cite{bian03,garcia05,garcia06}. 
The numerically observed structure of a seeded pressure blob with Gaussian shape is plume like, which is in accordance with cold ion plasma experiments~\cite{katz08}.
The relaxation of the thin-layer approximation allowed the treatment of realistic fluctuation levels and supported the results
of former studies~\cite{yu06,bodi08,angus14,omotani15}.
However, in hot ion plasmas  spatially compact blobs are observed experimentally~\cite{zweben02,boedo03,terry05,zweben15}. 
This blob structure can be ascribed to the finite Larmor radius (FLR) of the ions, which was demonstrated numerically by isothermal gyrofluid models 
for small~\cite{madsen11} and high fluctuation amplitudes~\cite{wiesenberger14}. 
For the sheath connected regime this was confirmed by reduced drift-fluid models~\cite{bisai12,bisai13,russel15}.
\\
Blobs are localised pressure perturbations, which in general involve particle density 
and temperature variations. 
Due to the relatively high ion to electron background temperature ratios $\tau_i$, ion temperature
 perturbations are in particular expected to have a profound influence on blob dynamics. 
Consequently we extend the former isothermal gyrofluid work~\cite{madsen11,wiesenberger14} 
to include ion and electron temperature dynamics and dynamic FLR effects, which account for 
variations in magnetic field and temperature. 
We highlight differences in blob propagation, particle transport and scaling to the isothermal model. 
Our model does not adopt the thin-layer approximation and is energetically consistent. 
Parallel dynamics and sheath effects have been studied by means of three dimensional 
numerical simulations~\cite{angus12,walkden13,easy14,halpern14,kendl15,walkden15} but are neglected here. 
Thus, our work is relevant for blobs in the inertial regime. 
\\
Our numerical investigation reports that finite temperature variations increase the maximal radial and total centre of mass blob velocity. 
Finite ion temperature results in a compact blob structure with significant poloidal velocity.
In an extensive parameter study we find that the maximal radial centre of mass 
velocities $\textrm{max}(V_{c,x})$ are in excellent agreement with the 
inertial velocity scaling law 
$V_{\perp} \equiv\sqrt{\frac{\sigma (\Delta p_{e}+\Delta p_{i})}{2 R n_{e0} m_i}}$. Here, $\Delta p_{e}$ and  $\Delta p_{i}$ are the electron and ion pressure amplitudes, $R$ is the radial distance to the outboard mid-plane, $n_{e0}$ is
the background electron density  and $m_i$ is the ion mass.
We quantify the FLR effect strength by the ratio of the magnetic field aligned component of the ion diamagnetic vorticity $\Omega_{d0}\equiv \vec{\hat{b}} \cdot \vec{\nabla} \times \vec{u}_d$ to 
the magnetic field aligned component of the $\vec{E}\times\vec{B}$ vorticity $\Omega_{E0}\equiv \vec{\hat{b}} \cdot \vec{\nabla} \times \vec{u}_E\sim V_\perp/\sigma$ 
to show that a smooth transition to a compact blob regime occurs for a ratio of unity. 
With the help of the FLR strength parameter $\Theta \equiv \sqrt{\frac{\Delta p_{i}^2 }{\left(\Delta p_{e}+\Delta p_{i}\right) t_{e0}n_{e0}}\frac{2 R \rho_{s0}^2}{\sigma^3 }}$ an empirical poloidal and total velocity scaling law 
is deduced and verified. Here, $\rho_{s0}=\sqrt{m_i t_{e0}}/(e B_0)$ is the drift scale with reference magnetic field magnitude $B_0$ and ion particle charge $e$.
\\
The remainder of the manuscript is organised as follows. We start with the derivation of the gyrofluid model equations in~\ref{sec:model}. In this section we derive also an
energy theorem and the vorticity density equation with its associated scaling laws. The described thermal gyrofluid model is solved numerically with seeded blobs as initial condition.
The numerical results of our parameter study are presented in~\ref{sec:simulation}. Here, we lay the focus of the discussion on thermal effects on propagation, compactness and particle transport 
of the blob. The results are discussed and summarised in~\ref{sec:conclusion}.
\section{Gyrofluid model}\label{sec:model}
Our approach relies on a full-F gyrofluid model~\cite{madsen13} for subsonic $\vec{E}\times\vec{B}$ flows, 
which emerges by taking the gyrofluid moments over the gyrokinetic Vlasov-Maxwell equations~\cite{brizard07}. 
This reduces the dimensionality and consequently the computational cost drastically.
No distinction is made between the dynamical background and the low-frequency fluctuations. 
The ratio of gradient length-scale to the ion gyro-radius $\rho_i$ may approach unity
but polarisation effects are taken in the long wavelength limit (LWL) $k_\perp^4 \rho_i^4 \ll 1$.
Finite Larmor radius (FLR) effects arise naturally in gyrofluid models resulting in a rather simple set of equations. 
FLR-corrected drift-fluid models are recovered when the gyrofluid model is taken in the LWL~\cite{scott07}.
\\
In the following we derive a four field full-F gyrofluid model for an electrostatic magnetised plasma.
The model is derived for a 2D slab geometry, where the magnetic field is varying in the radial direction and is perpendicular
to the 2D slab.
Collisional terms and parallel dynamics are discarded whereas we keep all nonlinearities in order to obey the energy theorem 
and to get a proper description of the dynamics even for high fluctuation amplitudes.
This results in equations for a gyro-centre  density $N$ and perpendicular gyro-centre temperature $T$, 
which are coupled via the nonlinear polarisation equation. 
Due to the neglect of parallel dynamics our gyrofluid model applies to the inertial regime.
\subsection{Polarisation equation}\label{sec:polarisation}
\begin{figure}[!ht]
\centering
\includegraphics[trim = 0px 0px 0px 0px, clip, scale=0.8]{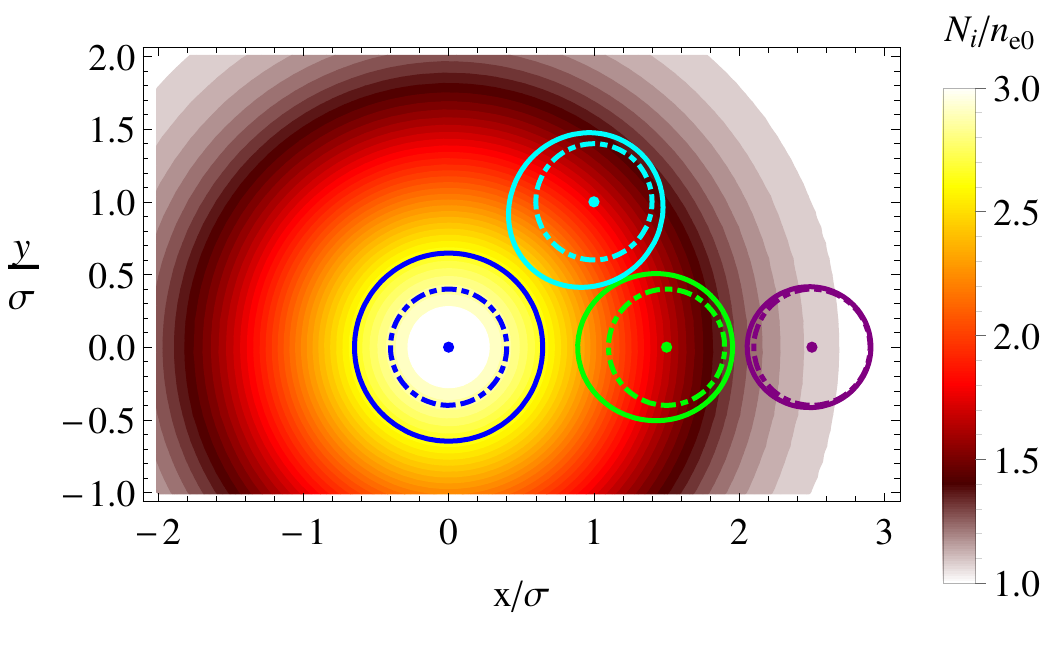}
\caption{The set of gyro-centres is shown with \(|\vec{x}-\vec{r}|^2 = \rho_i(\vec{x})^2\) for four different positions $\vec{r}$ in case of a Gaussian ion gyro-centre pressure blob. 
         The contours illustrate the normalised ion gyro-centre  density $N_i/n_{e0}$ or ion gyro-centre temperature $T_i/t_{i0}$ for an ion to electron background temperature ratio $\tau_i=4$, cross-field size $\sigma=5$ and amplitude $A=2$ (cf.~\ref{sec:initialisation}). The curve for 
         a constant gyro-radius (dot-dashed contours) is always a circle with radius $\rho_i (T_{i0},B_0) $. On the contrary the curve for a 
         dynamic gyro-radius (solid contours) is either a circle with radius $ \rho_i(T_i,B) \geq \rho_i(T_{i0},B_0)$ or a deformed circle. }
\label{fig:gyroavg}
\end{figure}

The polarisation equation is the gyrofluid expression of quasi neutrality and reads
\begin{align}\label{eq:polarisationeq}
  n_e -\Gamma_{1,i}^\dagger N_i = \vec{\nabla} \cdot\left(\frac{N_i}{\Omega_i B} \vec{\nabla}_\perp \phi\right),
\end{align}
where $n_e$ is the electron density, $N_i$ is the ion gyro-centre density, $\phi$ 
is the electrostatic potential and $\Omega_i=e B/m_i$ is the ion gyro-frequency. 
We note here that as a consequence of the gyro-centre transformation~\cite{brizard07} the ion gyro-centre density $N_i$ 
is not to be confused with the ion particle density $n_i\neq N_i$.
The perpendicular gradient is defined by 
$\vec{\nabla}_\perp \equiv -\vec{\hat{b}} \times (\vec{\hat{b}} \times \vec{\nabla})$ and the unit vector in the magnetic field direction by $\vec{\hat{b}}\equiv \vec{B}/B$.
The mass ratio between the electrons and ions is small, permitting us to neglect electron FLR and finite electron inertia effects. 
This casts the electron gyro-centre variables into common fluid variables $\left(N_e,T_{e}\right) \equiv \left(n_e,t_{e}\right)$.  
The sum of the electron density $n_e$ and the FLR corrected ion gyro-centre density $N_i$ is referred to as the nonlinear polarisation charge density (right hand side of~\eqref{eq:polarisationeq}).
In the polarisation~\eqref{eq:polarisationeq} no thin-layer approximation 
$\vec{\nabla} \cdot\left(\frac{N_i}{\Omega_i B} \vec{\nabla}_\perp \phi\right) \approx \frac{N_{i0}}{\Omega_{i0} B_0} \vec{\nabla}_\perp^2 \phi$ 
is applied and the full nonlinear polarisation charge density is retained. 
Here, and in all other abbreviations the subscript '0' denotes a constant background quantity.
We note that FLR corrections to the polarisation density are included in delta-f gyrokinetic and gyrofluid models~\cite{madsen11}, 
but are omitted in full-F gyrokinetic as well as full-F gyrofluid models. Therefore, even if we invoked the thin-layer approximation to 
the full-F model, it would only agree with delta-f gyrofluid models in the LWL~\cite{wiesenberger14}. 
In gyrofluid models the ion polarisation drift enters via the nonlinear polarisation charge density. 
The gyroaveraging operators appear as the Pad\'e approximants~\cite{dorland93}
\begin{align}
 \Gamma_{1,i} = \frac{1}{1-\frac{\rho_i^2}{2} \vec{\nabla}_\perp^2}, \\
 \Gamma_{1,i}^\dagger = \frac{1}{1- \vec{\nabla}_\perp^2\frac{\rho_i^2}{2}}.
\end{align}
Note here the temperature and magnetic field dependence of the ion gyro-radius $\rho_i = \sqrt{T_{i}/ m_i}/\Omega_i$, which introduces dynamic FLR effects $\rho_i(T_i,B)$.
In contrast to isothermal gyrofluid models with constant FLR effects $\rho_i(T_{i0},B_0)$~\cite{wiesenberger14} the $\Gamma_{1,i} $ operator is no longer self-adjoint $\Gamma_{1,i} \neq \Gamma_{1,i}^\dagger$ and the 
$\Gamma_{2,i} \equiv \frac{\rho_i}{2} \frac{\partial \Gamma_{1,i} }{\partial \rho_i}   $ operator
appears in the gyrofluid moment equations due to spatial variations in the temperature or magnetic field
\begin{align}
 \Gamma_{2,i} = \frac{\frac{\rho_i^2}{2} \vec{\nabla}_\perp^2}{\left(1-\frac{\rho_i^2}{2} \vec{\nabla}_\perp^2\right)^2},\\
 \Gamma_{2,i}^\dagger = \frac{\vec{\nabla}_\perp^2\frac{\rho_i^2}{2} }{\left(1- \vec{\nabla}_\perp^2\frac{\rho_i^2}{2}\right)^2}.
\end{align}
The FLR corrected ion gyro-centre density $\Gamma_1^\dagger N_i$ in the polarisation~\eqref{eq:polarisationeq} 
is the averaged charge contribution at a position $\vec{r}$ from gyro-centres whose gyro-orbits intersect $\vec{r}$.
In case of constant FLR effects the mean gyro-orbit is constant and these gyro-centres are placed on a circle with constant radius $\rho_i(T_{i0},B_0)$ at position $\vec{r}$.
On the contrary, for dynamic FLR effects the mean gyro-orbit is spatially varying, so that these gyro-centres are placed on an arbitrary curve. 
For the simple case of a Gaussian ion gyro-centre pressure blob this curve reduces to a circle with radius $\rho_i(T_{i},B)$ or a deformed circle at position $\vec{r}$, which is shown in~\ref{fig:gyroavg}.
As a consequence the curves sample different ion gyro-centre  densities for constant or dynamic FLR effects.
This, in comparison to constant FLR effects, effectively increases or decreases the polarisation charge density in the blob centre or edge respectively. 
\\
Due to the neglect of electron FLR effects the gyroaveraging operators for electrons reduce to $\Gamma_{1,e}=\Gamma_{1,e}^\dagger=1$ and $\Gamma_{2,e}=\Gamma_{2,e}^\dagger=0$. 
\subsection{Gyrofluid moment equations}\label{sec:gyromoment}
The time evolution of a gyro-centre density $N$ is governed by the zeroth gyrofluid moment over the gyrokinetic Vlasov equation
\begin{align}\label{eq:zerothgyromoment}
 \frac{\partial}{\partial t} N +
\vec{\nabla} \cdot \left(N \left[\vec{U}_E  + \vec{U}_\eta + \vec{U}_{\vec{\nabla} B} \right]
\right)
=  \Lambda_N .
\end{align}
Here, the gyrofluid drifts (denoted with capital $U$) account for the 
$\vec{E}\times\vec{B}$ drive by the $\vec{E}\times\vec{B}$ drift $\vec{U}_E $, the interchange drive by the ${\vec{\nabla} B}$ drift $ \vec{U}_{\vec{\nabla} B}$
and FLR corrections to the gyro-averaged electric potential due to variations in the temperature or magnetic field by  $\vec{U}_\eta$:
\begin{align}\label{eq:UE}
 \vec{U}_E \equiv\frac{  \vec{\hat{b}} \times  \vec{\nabla} \psi }{B}, \\
 \label{eq:UgradB}
   \vec{U}_{\vec{\nabla} B} \equiv\frac{T \vec{\hat{b}} \times \vec{\nabla}\ln B }{q B},  \\
   \label{eq:Ueta}
 \vec{U}_\eta \equiv\frac{ \left(\Gamma_{2} \phi \right) \vec{\hat{b}} \times  \vec{\nabla} \eta }{B}.
\end{align}
Note that as a consequence of the choice of our slab geometry, which is introduced later in this section, no curvature drift appears in the gyrofluid moment equations.
In ~\eqref{eq:UE} and~\eqref{eq:Ueta}  we introduced the gradient of temperature and magnetic field $ \vec{\nabla} \eta  $, the generalised potential $ \psi $ and the $\vec{E}\times\vec{B}$ 
drift velocity $ \vec{u}_E $:
\begin{align} \label{eq:gradeta}
 \vec{\nabla} \eta = \vec{\nabla} \ln{B} - \vec{\nabla} \ln{T}, \\ \label{eq:psi}
 \psi  = \Gamma_{1} \phi - \frac{m u_E^2 }{2 q}, \\ \label{eq:uE}
 \vec{u}_E = \frac{1}{B} \vec{\hat{b}} \times \vec{\nabla} \phi .
\end{align}
Here, $q$ is the particle charge. 
Since the gyrokinetic Lagrangian is taken in the LWL, polarisation effects are retained in the generalised potential $\psi$ of~\eqref{eq:psi} only via the
$\vec{E}\times \vec{B}$ energy term~\cite{madsen13}.
The dissipative term in the gyrofluid moment~\eqref{eq:zerothgyromoment} is a  hyperdiffusive term of second order 
$ \Lambda_{N}\equiv  -\nu \nabla_\perp^4 N $.
The second gyrofluid moment yields the partial differential equation for a perpendicular gyro-centre pressure $P$. The equation is rewritten with the help
of the zeroth gyrofluid moment~\eqref{eq:zerothgyromoment} into an evolution equation for a perpendicular gyro-centre temperature $T$:
\begin{align}\label{eq:secgyromoment}
  \frac{\partial }{\partial t}  T     &+T   \vec{\nabla}  \cdot \left(  \vec{U}_{E,2}  + \vec{U}_{\vec{\nabla} B} +  \vec{U}_{\eta}\right) \nonumber\\ &
 +T \left( \vec{U}_{E,2}  + \vec{U}_{\vec{\nabla} B} +  \vec{U}_{\eta} \right) \cdot \vec{\nabla} \ln N \nonumber\\ &
 - T  \left(\vec{U}_{E} + \vec{U}_{E,2}  + 2 \vec{U}_{\vec{\nabla} B}  \right) \cdot \vec{\nabla} \eta  
 =  \Lambda_{T}. \nonumber \\
\end{align}
Here, the gyrofluid drift $\vec{U}_{E,2}$ captures FLR corrections to the $\vec{E}\times\vec{B}$ drift $\vec{U}_E $, which emerge due 
to spatial variations in the gyro-centre temperature $T$ or the magnetic field $B$:
\begin{align}
 \label{eq:UE2}
  \vec{U}_{E,2} \equiv \frac{\vec{\hat{b}} \times  \vec{\nabla}  \Gamma_{2} \phi }{B}.
\end{align}
In the second gyrofluid moment~\eqref{eq:secgyromoment} a hyperdiffusive term of second order $ \Lambda_{T}\equiv -\nu \nabla_\perp^4 T$ is added, which together with $\Lambda_N$ ensures numerical stability.
\\
Our gyrofluid model employs right-handed Cartesian coordinates $(x,y,z)$ and a slab magnetic field $\vec{\hat{b}} = \vec{\hat{e}}_z$ with a radially varying magnitude $ \frac{1}{B}= \frac{1}{B_0} \left( 1+x/R\right) $. 
Here, $R$ is the radial distance to the outboard mid-plane and $B_0$ is the reference magnetic field magnitude.
In this simple geometry a more accessible representation of the electron and ion gyrofluid moment~\eqref{eq:zerothgyromoment} and 
~\eqref{eq:secgyromoment} is obtained by expressing the set of equations with the help of Poisson brackets and curvature operators. 
The slab geometry approximation reduces the curvature operator to  $\mathcal{K} (f) \equiv \nabla \cdot \left(\frac{1}{B}\vec{\hat{b}} \times\vec{\nabla} f \right) =-\frac{1}{B_0 R } \frac{\partial f}{\partial y }$
and the Poisson bracket to $ \left[f,g \right]_{x y} = \frac{\partial f}{\partial x } \frac{\partial g}{\partial y } -\frac{\partial g}{\partial x } \frac{\partial f}{\partial y }$.
We rearrange the terms in the gyrofluid moment~\eqref{eq:zerothgyromoment} and 
~\eqref{eq:secgyromoment} to  highlight the highly nonlinear nature of our system:
\begin{align}
\label{eq:dtne}
\frac{\partial}{\partial t} n_e =&-
\frac{1}{B} \left[\phi,n_e \right]_{x y} - n_e \mathcal{K}\left(\phi \right) + \frac{1}{e} \mathcal{K}\left(t_{e} n_e \right) 
\nonumber \\ 
&+  \Lambda_{n_e}, 
 \\
\label{eq:dtNi}
\frac{\partial}{\partial t} N_i =&
-\frac{1}{B} \left[\psi_i,N_i \right]_{x y} 
+\frac{1}{B} \left[\ln T_{i},N_i \chi_i \right]_{x y} \nonumber \\ 
&- N_i \mathcal{K}\left(\psi_i + \chi_i\right) 
+ N_i \chi_i \mathcal{K}\left(\ln T_{i} -  \ln N_i \right) \nonumber \\ 
&
- \frac{1}{e} \mathcal{K}\left(T_{i} N_i \right) 
+  \Lambda_{N_i} , \\
\label{eq:dtte}
   \frac{\partial }{\partial t}  t_{e} =&
   -\frac{1}{B} \left[ \phi , t_{e} \right]_{x y}  
 -t_{e}   \mathcal{K} (\phi )  
+ \frac{3 t_{e}}{ e }   \mathcal{K} (t_{e})\nonumber \\ 
&
+  \left(\frac{t_{e}^2}{e}  \right)\mathcal{K} \left(  \ln  n_e  \right) 
 + \Lambda_{t_{e} },  \\
\label{eq:dtTi}
    \frac{\partial }{\partial t}  T_{i} =&
-\frac{1}{B} \left[ \psi_i + 2\chi_i ,T_{i} \right]_{x y}  \nonumber \\
&- \frac{T_{i} \chi_i}{B } \left[\ln  \chi_i - \ln T_{i} , \ln N_i\right]_{x y}
 \nonumber \\ &
 -T_{i}  \mathcal{K} (\psi_i + 3\chi_i)  
- \left(\frac{3 T_{i} }{e}   - \chi_i \right)\mathcal{K} (T_{i} )\nonumber \\ 
&
-  \left(\frac{T_{i} ^2}{e} + T_{i}   \chi_i \right)\mathcal{K} \left(  \ln  N_i  \right) + \Lambda_{T_{i} }.
\end{align}
Here, we introduced $\chi_i \equiv \Gamma_{2,i} \phi$.
In the isothermal limit the temperatures are constant $\left(t_{e},T_{i}\right) = \left(t_{e0},T_{i0}\right)$ and second order FLR effects vanish so that 
$ \Gamma_{1,i} = \Gamma_{1,i}^\dagger$ and $ \Gamma_{2,i} = \Gamma_{2,i}^\dagger  = 0$.
From this we obtain the set of full-F gyrofluid equations, which was studied in~\cite{wiesenberger14}.
\subsection{Energy theorem}\label{sec:energytheorem}
The energy $\mathcal{E} $ of our system can be determined by integrating the sum of the zeroth 
gyrofluid moment equation multiplied by the factor $(T + q \psi)$ and the second gyrofluid moment equation over the domain. 
We obtain the energy theorem $ \frac{\partial}{\partial t}\mathcal{E} =\Lambda$ for electrons and ions by neglecting boundary terms:
\begin{align}\label{eq:energytheorem}
 \mathcal{E} =& \int d\vec{x} \left(n_e t_{e} +N_i T_{i} + \frac{m_i N_i u_E^2}{2} \right) , \\
\Lambda = &
 \int d\vec{x}  \bigg[\left(t_{e} - e \phi \right) \Lambda_{n_e}  +\left(T_{i} + e \psi_i \right) \Lambda_{N_i} \nonumber \\ & +
  n_e  \Lambda_{t_{e}}+ 
 \left(1+\frac{e}{T_{i}} \chi_i\right)N_i  \Lambda_{T_{i}}\bigg].
\end{align}
The energy consists of the internal (thermal) energy densities for electrons and ions and the $\vec{E}\times\vec{B}$ energy density. 
The dissipative terms $\Lambda_N,\Lambda_T$ enter the energy theorem via $\Lambda$. 
The derived energy theorem resembles those in~\cite{madsen13,madsen15}.
\subsection{Vorticity density equation}\label{sec:vorticityequation}
In order to understand the relation between gyro-centre and particle fields and 
to show the correspondence to drift-fluid models explicitly we derive 
the inherent vorticity density equation of the presented full-F gyrofluid model in the LWL~\cite{scott07}.
Thus we first invert the polarisation~\eqref{eq:polarisationeq} and apply the LWL. This yields 
an expression for $N_i$, which depends on the physical meaningful variables $n_e$, $t_{i}$ and $\phi$~\cite{madsen15}:
\begin{align} \label{eq:iongyrocenterdensity}
 N_i &\approx n_e - \vec{\nabla}_\perp^2\left(\frac{n_e t_{i}}{2 m_i \Omega_i^2}\right) - \vec{\nabla} \cdot \left( \frac{n_e}{B \Omega_i}\vec{\nabla}_\perp \phi\right) \nonumber \\ 
 &\equiv n_e\left(1-\frac{\omega^* }{\Omega_{i}}\right).
\end{align}
Here, we defined the sum of the $\vec{E}\times\vec{B}$  vorticity plus half of the ion diamagnetic vorticity as
\begin{align} \label{eq:omegastar}
 \omega^* &= \frac{\Omega_{i}}{n_{e}}\left[
  \vec{\nabla} \cdot \left( \frac{n_e}{B \Omega_i}\vec{\nabla}_\perp \phi\right) +
  \vec{\nabla}_\perp^2\left(\frac{n_e t_{i}}{2 m_i \Omega_i^2}\right)\right]\nonumber \\
  &\equiv  \Omega_E + \frac{1}{2}\Omega_d.
\end{align}
The generalised vorticity $\Omega^*= \omega^* +\frac{1}{2}\Omega_d $ is the sum of the magnetic field aligned component of the $\vec{E}\times\vec{B}$  vorticity and the ion diamagnetic vorticity. 
We note here that the $\vec{E}\times\vec{B}$ vorticity consists of the magnetic field aligned component of the common $\vec{E}\times\vec{B}$  vorticity
$ \Omega_{E0} \equiv \vec{\hat{b}} \cdot  \vec{\nabla} \times \vec{u}_E = \vec{\nabla} \cdot \left(\frac{1}{B} \vec{\nabla}_\perp \phi\right) $ plus a cross 
term $\Omega_{EX} \equiv \frac{1}{B} \vec{\nabla} \ln \left(\frac{n_e}{\Omega_i} \right)\cdot\vec{\nabla}_\perp \phi$. Analogously the magnetic field aligned component of the ion diamagnetic vorticity consists of 
$\Omega_{d0} \equiv  \vec{\hat{b}} \cdot \vec{\nabla} \times \vec{u}_d =  \vec{\nabla} \cdot \left(\frac{1}{\Omega_i m_i n_e} \vec{\nabla}_\perp p_{i\perp}\right) $
and a cross term $ \Omega_{dX} \equiv \frac{ t_{i\perp}}{m_i \Omega_i  }\big[ 4 (\vec{\nabla}_\perp \ln \Omega_i)^2 -2 \vec{\nabla}_\perp^2 \ln \Omega_i- 
              \vec{\nabla}_\perp \ln p_{i\perp} \cdot \left(3 \vec{\nabla}_\perp\ln \Omega_i + \vec{\nabla}_\perp\ln n_e\right)\big]$.
The generalised vorticity density is given by
$ \mathcal{W}\equiv n_{e}\Omega^* $. 
The LWL vorticity density equation is derived by taking the material derivative $\frac{d f}{dt}\equiv \frac{\partial f}{\partial t} + \frac{1}{B}\left[\phi, f\right]_{xy}$ over the generalised vorticity density
\begin{align}\label{eq:voreq}
\frac{1}{\Omega_i }\frac{d }{d t}\mathcal{W}  \approx& \frac{1}{e} \mathcal{K} \left(p_{e} + p_{i}    \right) 
 -  \frac{1}{B} \left[\vec{\nabla}_\perp\phi,\vec{\nabla}_\perp \left( \rho_i^2 n_e \right) \right]_{x y} 
 \nonumber \\
  &
  +\frac{1}{\Omega_i } \Lambda_{\mathcal{W}},
\end{align}
where we have neglected FLR corrections in curvature operators and nonlinear terms that fall into the LWL. Dynamic FLR effects arise from the ion diamagnetic contribution of the generalised vorticity density $\mathcal{W}$ and 
the second term on the right hand side of the vorticity density~\eqref{eq:voreq}.
The dissipation of the generalised vorticity density is represented by $\Lambda_{\mathcal{W}}$, which is discussed in detail in
\ref{sec:dissipation}. 
We note here that the Prandtl number $\mathit{Pr}\equiv \frac{\nu_{\Omega^*}}{\nu_T}$
and Schmidt number $\mathit{Sc}\equiv \frac{\nu_{\Omega^*}}{\nu_N}$ are fixed to unity. Hence, the mass diffusivity $\nu_N$, 
the thermal diffusivity $\nu_T$ and the kinematic viscosity $\nu_{\Omega^*}$ reduce to a single dissipative parameter $\nu$. 
In the absence of dissipative terms we can rewrite the vorticity density~\eqref{eq:voreq} to a form, which equals the divergence of the ion polarisation flux. It reads
\begin{align}
  \vec{\nabla} \cdot \left(\frac{n_e}{B \Omega_i} \frac{d}{d t }\vec{\nabla}_\perp \phi^*\right) 
     &\approx
  \frac{1}{e}\mathcal{K}\left( p_{e} +  p_{i}   \right), 
\end{align}
and allows a comparison to FLR-corrected drift-fluid models~\cite{smolyakov98,madsen16}.  
Here, we defined $\vec{\nabla}_\perp \phi^*\equiv  \vec{\nabla}_\perp \phi  + \frac{t_i}{e} \vec{\nabla}_\perp \ln \left(\frac{p_i}{m_i \Omega_i^2}\right)$.
\subsection{Initialisation}\label{sec:initialisation}
Since gyro-centre fields $N_i, T_{i}$ match the particle fields $n_i,t_{i}$ only to zeroth order in $k_\perp^2 \rho_i^2$ and $e\phi/t_{i}$, 
an accurate initialisation procedure is required in order to avoid any residual $\vec{E}\times\vec{B}$ vorticity.
Hence, our initialisation routine relies on transformations from the gyro-centre fields 
$N_i$, $T_{i}$, $P_{i}= N_i T_{i}$ to the particle fields
$n_e=n_i$, $t_{i}$,  $t_{e}$, $p_{i} = n_i t_{i}$
without the generation of
artificial initial $\vec{E}\times\vec{B}$ vorticity. 
For the ion gyro-centre density and perpendicular temperature we mimic an initial blob by a Gaussian of the form
\begin{align}
 N_{i}\left(\vec{x},0\right) =   n_{e0}\left[1+A \exp{\left(-\frac{\left(\vec{x}-\vec{x}_0\right)^2}{2\sigma^2}\right)}\right],  \\
 T_{i}\left(\vec{x},0\right) =   t_{i0}\left[1+A \exp{\left(-\frac{\left(\vec{x}-\vec{x}_0\right)^2}{2\sigma^2}\right)} \right],
\end{align}
with initial amplitude $A$  of the gyro-centre fields  $N_{i}$ and $T_{i}$.
In order to initialise with zero $\vec{E}\times\vec{B}$ vorticity we take the polarisation~\eqref{eq:polarisationeq} and 
the equation for $p_{i}$~\cite{madsen15} in the limit $\phi = 0$. 
This yields the following initial particle fields
\begin{align}
 n_e  =\Gamma_{1,i}^\dagger N_i, \\
 p_{i} = \left(\Gamma_{1,i}^{\dagger} + \Gamma_{2,i}^{\dagger} \right)P_{i} .
\end{align}
The initial electron pressure is related to ion pressure via $p_{i} = \tau_i p_{e}$ so that the initial electron temperature is given by $t_{e} = t_{i}/\tau_i$.
We note here that the initial amplitudes and cross-field sizes of the particle fields $n_e$, $t_{e}$ and  $t_{i}$ may 
differ from the initial parameters $A$ and $\sigma$ of the gyro-centre fields $N_{i}$ and $T_{i}$ due to the manifestation of FLR effects. 
Hence, the true initial amplitudes 
$\Delta n_e \equiv \max{\left\{n_e\left(\vec{x},0\right)-n_{e0}| \vec{x} \in V\right\} }$, $\Delta t_{e} \equiv \max{\left\{t_{e}\left(\vec{x},0\right)-t_{e0} | \vec{x} \in V\right\} }$ 
and $\Delta t_{i} \equiv \max{\left\{\tau_i\left(t_{e}\left(\vec{x},0\right)-t_{e0} \right)| \vec{x} \in V\right\} }$ are 
taken into account for the derivation of quantities like the global interchange rate $\gamma_g$, which is defined in the next~\ref{sec:interchangedyn}. For the size of the blob we
take the initial parameter $\sigma$ of the gyro-centre fields. 
\subsection{Interchange dynamics}\label{sec:interchangedyn}
Now, we use the derived vorticity density~\eqref{eq:voreq} to obtain an estimate for the maximal perpendicular blob velocity $V_{\perp}$. 
For this sake we first employ the thin-layer approximation on~\eqref{eq:voreq} and neglect all terms, which are related to FLR effects or dissipation. 
Our scaling analysis relies on the ideal interchange rate $\gamma$ and the blob cross-field size $\sigma$ as typical time and spatial scale, which results in the 
typical velocity scale $V_\perp = \gamma\sigma$.
The characteristic scale of the generalised vorticity and perpendicular pressure is the ideal interchange rate $\gamma$
and the total scalar pressure perturbation $\Delta p_{e} + \Delta p_{i}$.
Following~\cite{manz13} we obtain the explicit expression for the ideal interchange rate $\gamma  =  \sqrt{\frac{\Delta p_{e}+\Delta p_{i}}{2 R n_{e0} \sigma m_i }}$.
We note that an initial electron and ion temperature perturbation $(\Delta t_e, \Delta t_i)$ enters the initial electron and ion pressure perturbation according to
$(\Delta p_{e} ,\Delta p_{i}) \sim (n_{e0}\Delta t_{e}+t_{e0}\Delta n_e+\Delta n_e \Delta t_{e},
n_{e0}\Delta t_{i}+t_{i0}\Delta n_e+\Delta n_e \Delta t_{i})$, which in the isothermal limit reduces to  $ (\Delta p_{e},\Delta p_{i})\sim(t_{e0}\Delta n_{e},t_{i0}\Delta n_{e})$. 
\\
Now, we postulate the factor $\sqrt{1+\Delta n_e/n_{e0}}$ `ex post' as a nonlinear correction factor to the blob size $\sigma$ and the ideal interchange rate $\gamma$.
As a result the typical spatial and time scale is the effective blob size $\sigma_g \equiv \sigma  \sqrt{1+\Delta n_e/n_{e0}}$ and the global interchange rate~\cite{wiesenberger14}
\begin{align}\label{eq:gammascalingthg}
 \gamma_g   =  \sqrt{\frac{\Delta p_{e}+\Delta p_{i}}{2 R (n_{e0} + \Delta n_e) \sigma m_i }} .
\end{align}
This form of scaling fits our simulation results best. 
With the help of the global interchange rate $\gamma_g=V_\perp/\sigma_g$ we obtain the scaling law for the maximal perpendicular blob velocity
\begin{align}\label{eq:velscalingth}
 V_{\perp} =\sqrt{\frac{\sigma (\Delta p_{e}+\Delta p_{i})}{2 R n_{e0} m_i}}.
\end{align}
The derived velocity scaling estimate of~\eqref{eq:velscalingth} differs for high electron density amplitudes $\Delta n_e$ 
from the so called global scaling $V_{\perp,g} \equiv  V_\perp/\sqrt{1+\Delta n_e/n_{e0}} $ 
as reported in~\cite{kube11,wiesenberger14}.
However, in~\cite{angus14,omotani15} the amplitude scaling of~\eqref{eq:velscalingth} was verified numerically for high fluctuation amplitudes in the cold ion temperature and isothermal limit.
In the isothermal limit our derived velocity scaling law agrees up to a factor $\sqrt{2}$ with the inertial velocity scaling law of~\cite{manz13}. 
This is reasoned in the low beta $\beta\equiv 2 p \mu_0/B^2\ll1$
approximation of the curvature operator. 
In this case the magnetic curvature is $\vec{\kappa} \equiv \vec{\hat{b}} \cdot \vec{\nabla} \vec{\hat{b}} \approx \vec{\nabla}_\perp \ln{B}$, 
whereas for a slab magnetic field the magnetic curvature is $\vec{\kappa} =0$. 
Note that \eqref{eq:velscalingth} also coincides with~\cite{garcia05} up to a factor $2$. 
\\
Now, we define a parameter, which is a measure of the FLR strength in the inertial regime.
In order to do so we take the ratio of the magnetic field aligned component of the ion diamagnetic vorticity $\Omega_{d0}$ to the 
estimate for the magnetic field aligned component of the $\vec{E}\times\vec{B}$ vorticity $\Omega_{E0}\sim V_\perp/\sigma$ in the inertial regime.
This defines the FLR strength parameter as $\Theta \equiv \frac{\Omega_{d0}}{\Omega_{E0}}$ or 
\begin{align}\label{eq:flrstrengthth}
 \Theta & \approx \sqrt{\frac{\Delta p_{i}^2 }{\left(\Delta p_{e}+\Delta p_{i}\right) t_{e0}n_{e0}}\frac{2 R  \rho_{s0}^2}{\sigma^3  }}.
\end{align}
We note here that in the FLR strength parameter $\Theta$ the ion diamagnetic vorticity $\Omega_{d0}$ stems from the LWL over the gyroaveraging operator $\Gamma_{1,i}^\dagger$. 
\section{Numerical experiments}\label{sec:simulation}
The gyrofluid model, which consists of Equations ~\eqref{eq:polarisationeq},~\eqref{eq:dtne},~\eqref{eq:dtNi},~\eqref{eq:dtte} and ~\eqref{eq:dtTi}, is numerically solved by the FELTOR library~\cite{feltor}.
This library relies on discontinuous Galerkin (dG) methods to discretise perpendicular~\cite{einkemmer14} and parallel spatial derivatives~\cite{held15}. 
DG methods are very versatile in the choice of the desired order
of accuracy and retain a high degree of parallelism in the resulting algorithm. FELTOR exploits this on shared as well as 
distributed memory systems and efficiently executes on CPUs and GPUs. 
This is required by the high computational cost of the underlying model.
\\
We resolve our box of size $l_x=l_y=40 \sigma$ by $P=3$ polynomial coefficients and $N_x=N_y=320$ grid cells to ensure converged simulations.
Boundary conditions of the fields are periodic in y and Dirichlet in x direction. We verified our code with the help of the  energy theorem of~\eqref{eq:energytheorem} 
and by performing convergence tests in the $L^2$-norm for all implemented numerical operators.
\\
Our dimensionless set of equations employs the Bohm normalisation with drift scale
$\rho_{s0}=\sqrt{m_i t_{e0}/(e B_0)}$, reference ion gyrofrequency
$\Omega_{i0}=e B_0/m_i$ and  cold ion acoustic speed 
$c_{s0} = \rho_{s0} \Omega_{i0}$
so that our variables are transformed to
$\vec{x}\leftarrow\vec{x}/\rho_{s0}$, 
$t\leftarrow t\Omega_{i0}$, 
$\vec{B}\leftarrow\vec{B}/B_0 $,
$t_{e}\leftarrow t_{e}/t_{e0} $, 
$T_{i}\leftarrow T_{i}/t_{i0}  $, 
$n_e \leftarrow n_e /n_{e0} $, 
$N_i\leftarrow N_i /n_{e0} $ and $\phi \leftarrow \phi e/t_{e0} $.
\\
The physical parameters match those of an exemplary SOL on the low field side of the ASDEX Upgrade (AUG) tokamak. 
Here, the electron temperature is $t_{e0} = 20eV$ and the magnetic field magnitude is $B_0 = 2T$, 
which determines the drift scale to $ \rho_{s0}= 0.32 mm $ for a deuterium plasma.
The remaining parameters are $R_0 =1.65 m = 5156 \rho_{s0}$, $a=0. 5m =1562 \rho_{s0}$, $R=R_0 +a = 2.15 m = 6719\rho_{s0}$ 
and enter the dimensionless equations via the curvature parameter $\kappa = \rho_{s0}/R \approx 0.00015$. 
The parameter range of the ion background temperature is $t_{i0} = \{0,0.1,0.5,1,2,4\} t_{e0}$ and 
the initial amplitude and width of the blob is $A=\{0.1,0.5,1,2\}$ and  $\sigma=\{5,10,20\}\rho_{s0}$. 
The ratio between the effective gravity to the dissipative forces is known as the Rayleigh number $ \mathcal{R}a $. For unity
Prandtl number we get $ \mathcal{R}a = \frac{\sigma^3}{m_i R \nu^2} \frac{ \Delta n_e \left(t_{e0}+t_{i0}\right) }{ n_{e0}} $ in the isothermal case and
$\mathcal{R}a = \frac{\sigma^3 }{m_i R \nu^2}\frac{ \left(\Delta p_{e}+\Delta p_{i}\right) }{ n_{e0}}$ if we include temperature dynamics ($d T / d t \neq 0$). We fix
the Rayleigh number to $\mathcal{R}a = 10^5 $ in all our simulations. This determines the viscosity to be in the turbulent high Reynolds number regime, where no impact on the
maximal velocity of the blob is expected~\cite{garcia05,kube12}.\\
In the following we start with the description of the nonlinear evolution of thermal blobs in~\ref{sec:blobmotion}. 
A detailed comparison of the propagation of isothermal and thermal blobs is given in~\ref{sec:prop}.
The impact of the temperature dynamics on the persistence of the initial blob structure is given afterwards in~\ref{sec:comp}.
In the end we focus on thermal effects on the particle transport of blobs (\ref{sec:trans}).
\begin{figure*}[!ht]
\centering
\includegraphics[trim = 0px 0px 0px 0px, clip, scale=0.35]{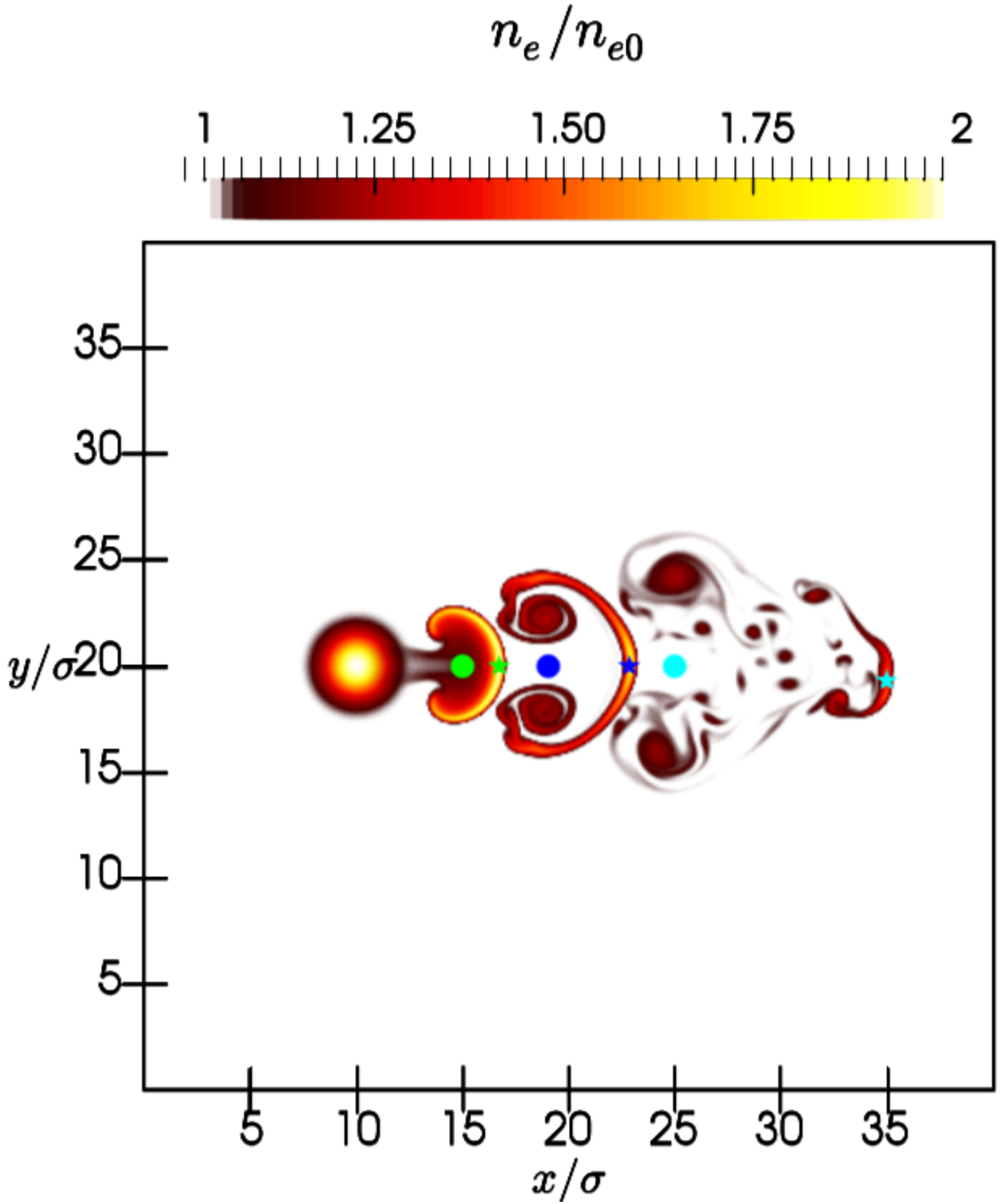}
\includegraphics[trim = 0px 0px 0px 0px, clip, scale=0.35]{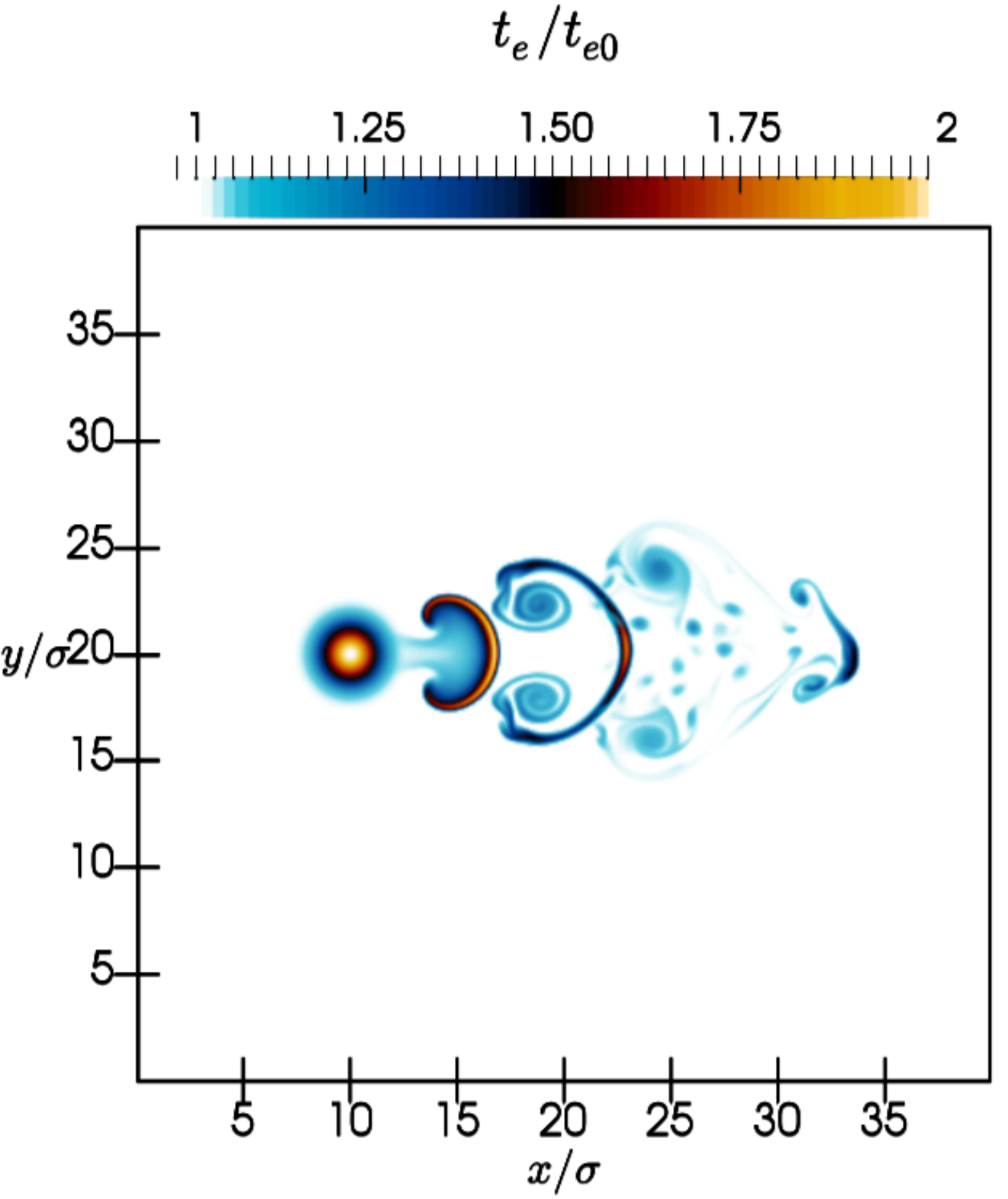}
\includegraphics[trim = 0px 0px 0px 0px, clip, scale=0.35]{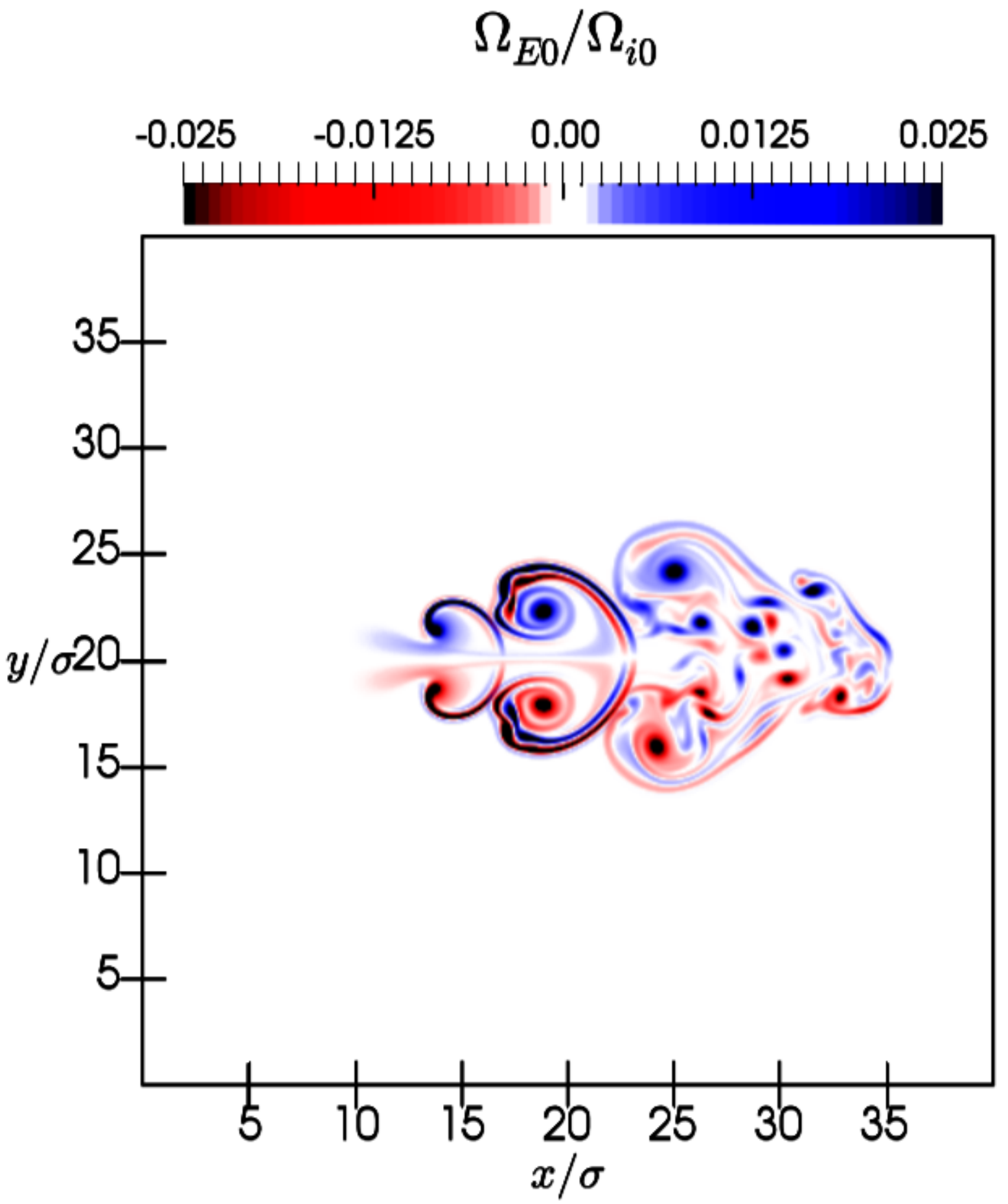}
\caption{The electron density $n_e$ (left), the electron temperature $t_{e}$ (centre) and the $\vec{E}\times\vec{B}$ vorticity $\Omega_{E0} \approx \frac{1}{B_0} \nabla_\perp^2 \phi$ (right)
is shown for a cold thermal blob with $\tau_i=0$, $A=1$, $\sigma=10$ at four different time steps $t= \left\{0,1300,2600,5000\right\} \Omega_{i0}^{-1}$. The COM position $\vec{R}_c$ (dot) and 
the position of the maximal electron density amplitude $\vec{R}_{max}$ (star) are visualised at three different time steps $t= \left\{1300,2600,5000\right\} \Omega_{i0}^{-1}$ (green, blue, cyan).
}
\label{fig:coldblob}
\end{figure*}
\begin{figure*}[!ht]
\centering
\includegraphics[trim = 0px 0px 0px 0px, clip, scale=0.34]{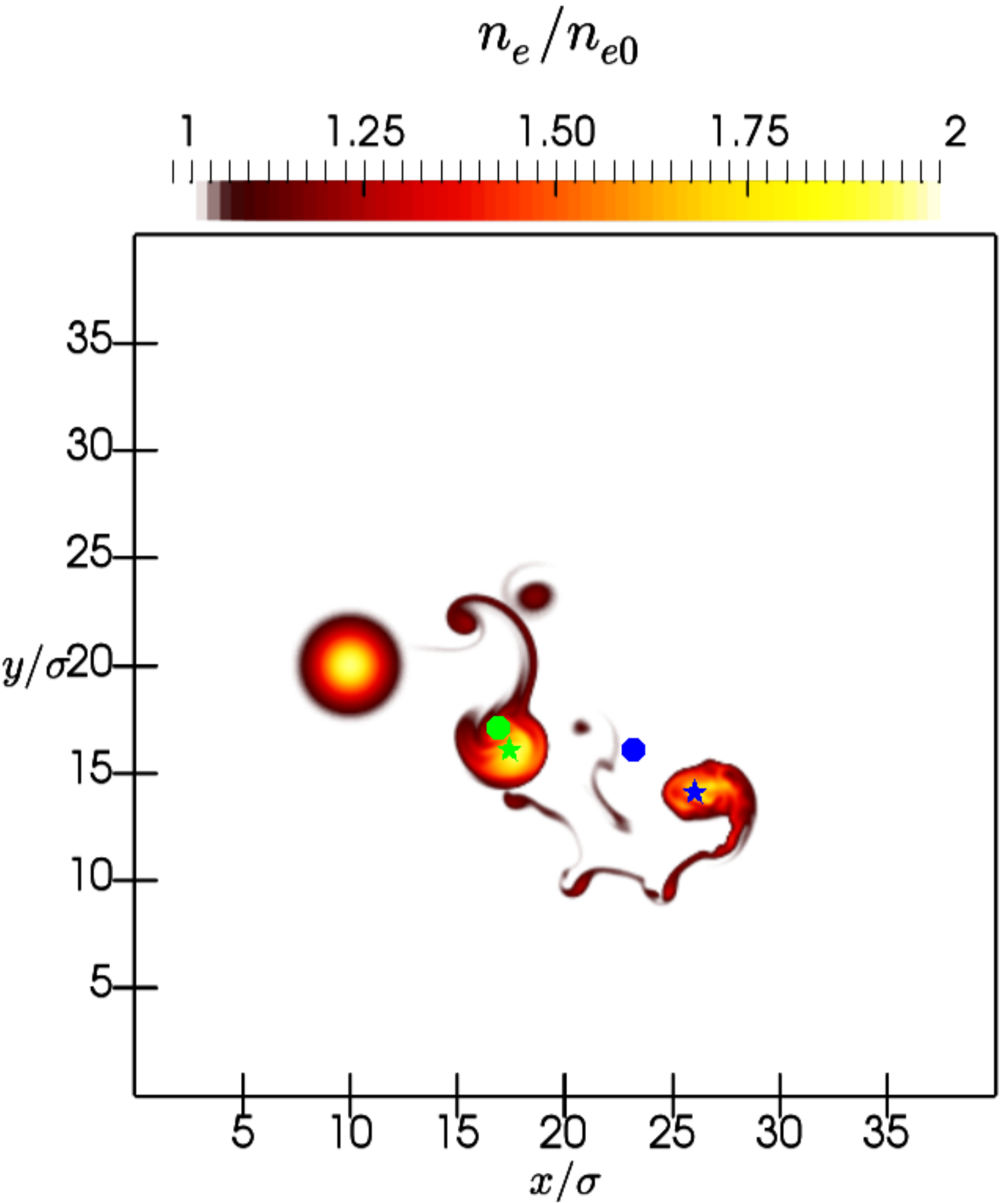}
\includegraphics[trim = 0px 0px 0px 0px, clip, scale=0.34]{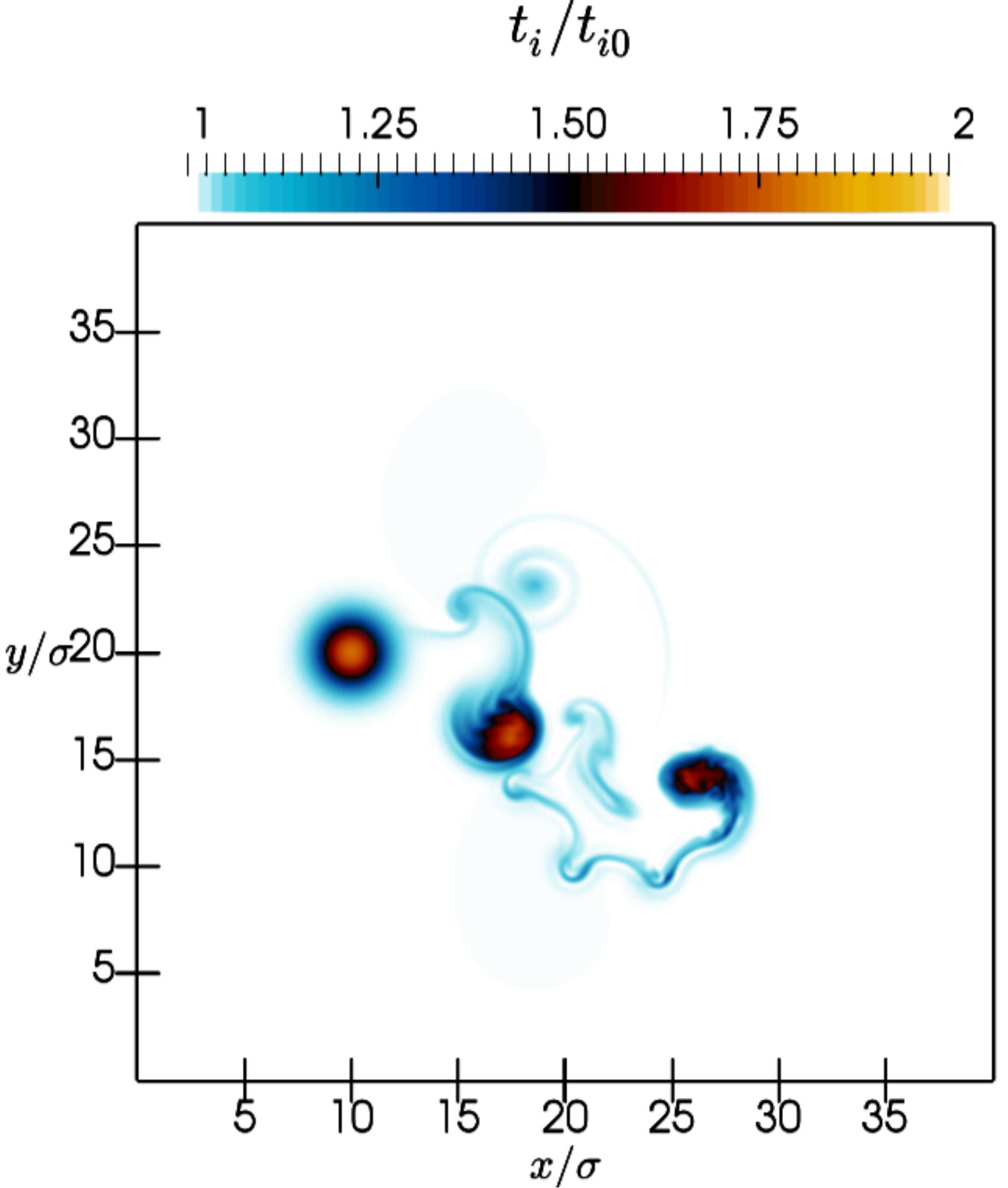}
\includegraphics[trim = 0px 0px 0px 0px, clip, scale=0.34]{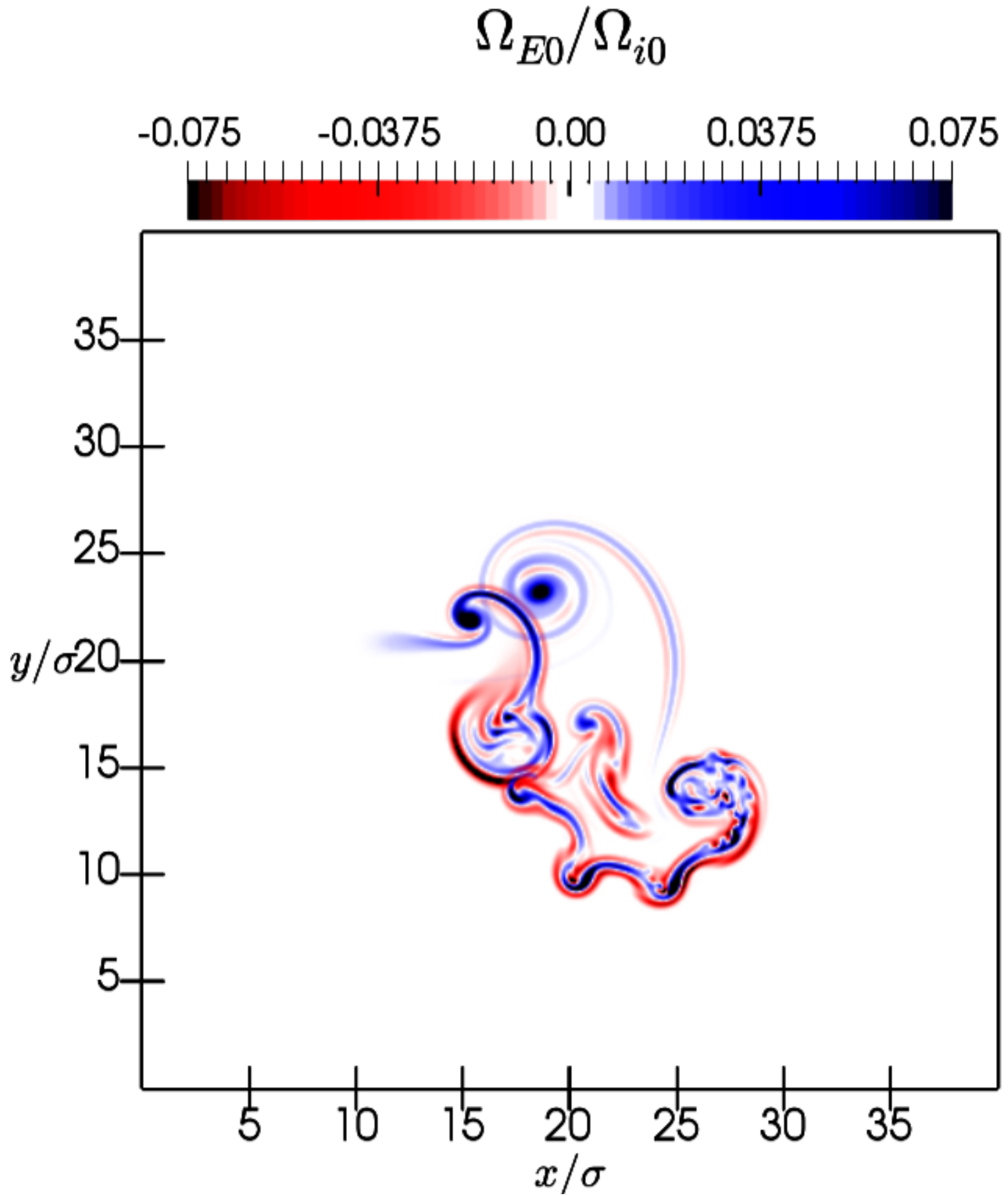}
\caption{The electron density $n_e$ (left), the ion temperature $t_{i}$ (centre) and the $\vec{E}\times\vec{B}$ vorticity $\Omega_{E0} \approx \frac{1}{B_0} \nabla_\perp^2 \phi$ (right)
is shown for a hot thermal blob with $\tau_i=4$, $A=1$, $\sigma=10$ at three different time steps $t= \left\{0,875,1875\right\}\Omega_{i0}^{-1}$. The COM position $\vec{R}_c$ (dot) and 
the position of the maximal electron density amplitude $\vec{R}_{max}$ (star) are visualised at two different time steps $t= \left\{875,1875\right\} \Omega_{i0}^{-1}$ (green, blue). }
\label{fig:hotblob}
\end{figure*}
\subsection{Nonlinear evolution of thermal blobs}\label{sec:blobmotion}
In the cold ion limit a finite electron pressure gradient induces an $\vec{E}\times\vec{B}$-vorticity dipole due to the $\vec{\nabla} B$ drift. This accelerates the blob into the radial direction~\cite{garcia06}. 
A steep and vertically extended pressure front develops finally into an up-down symmetric mushroom shape. This symmetry is broken in the late purely nonlinear phase due to the
up-down asymmetric nature of the gyrofluid model. We depict this behaviour for electron density $n_e$, electron temperature $t_e$ and the $\vec{E}\times\vec{B}$ vorticity $\Omega_{E0} \approx \frac{1}{B_0} \nabla_\perp^2 \phi$
of an exemplary cold blob in~\ref{fig:coldblob}.
\\

\begin{figure}[!ht]
\centering
\includegraphics[trim = 0px 0px 0px 0px, clip, scale=0.21]{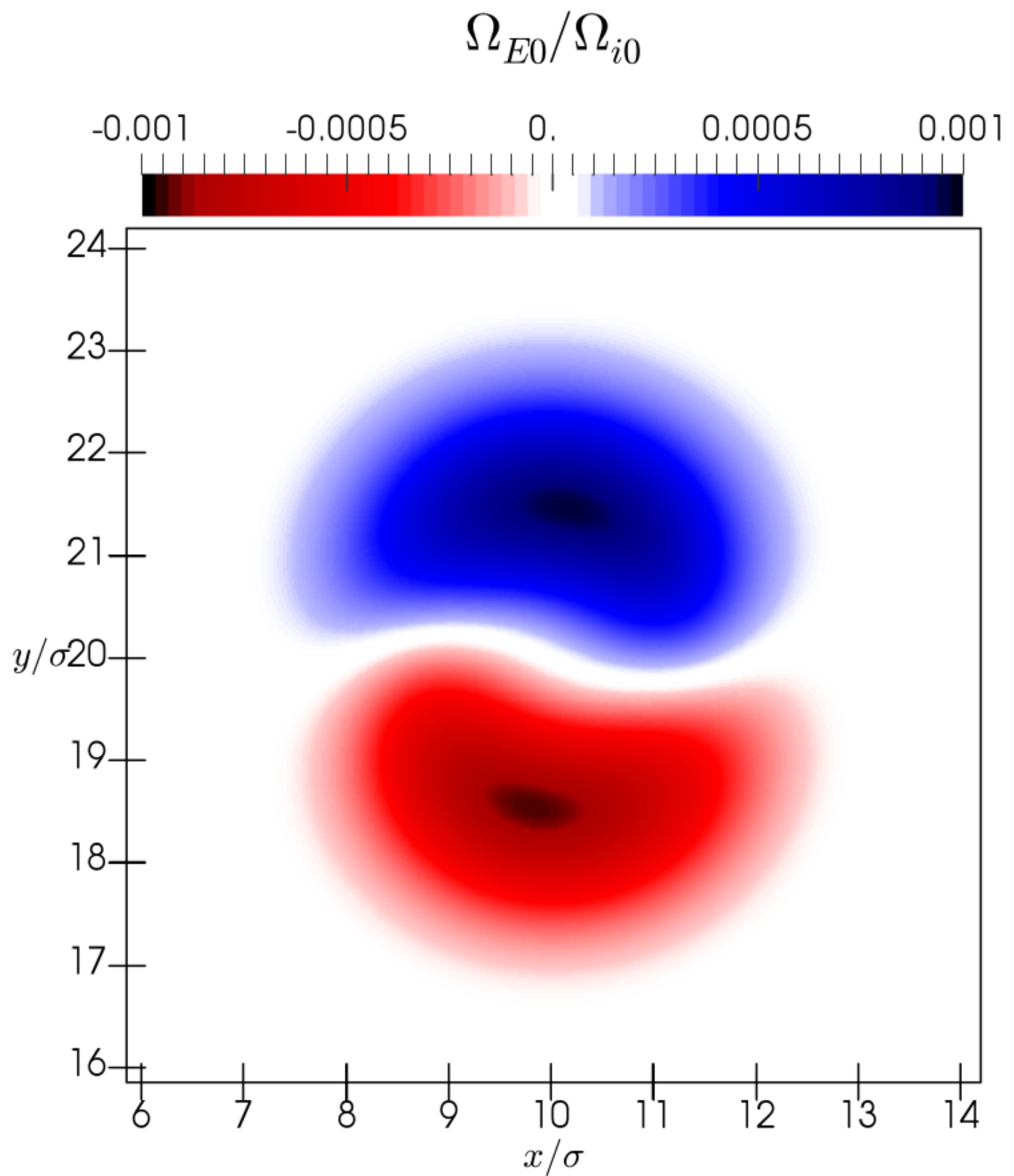}
\includegraphics[trim = 0px 0px 0px 0px, clip, scale=0.21]{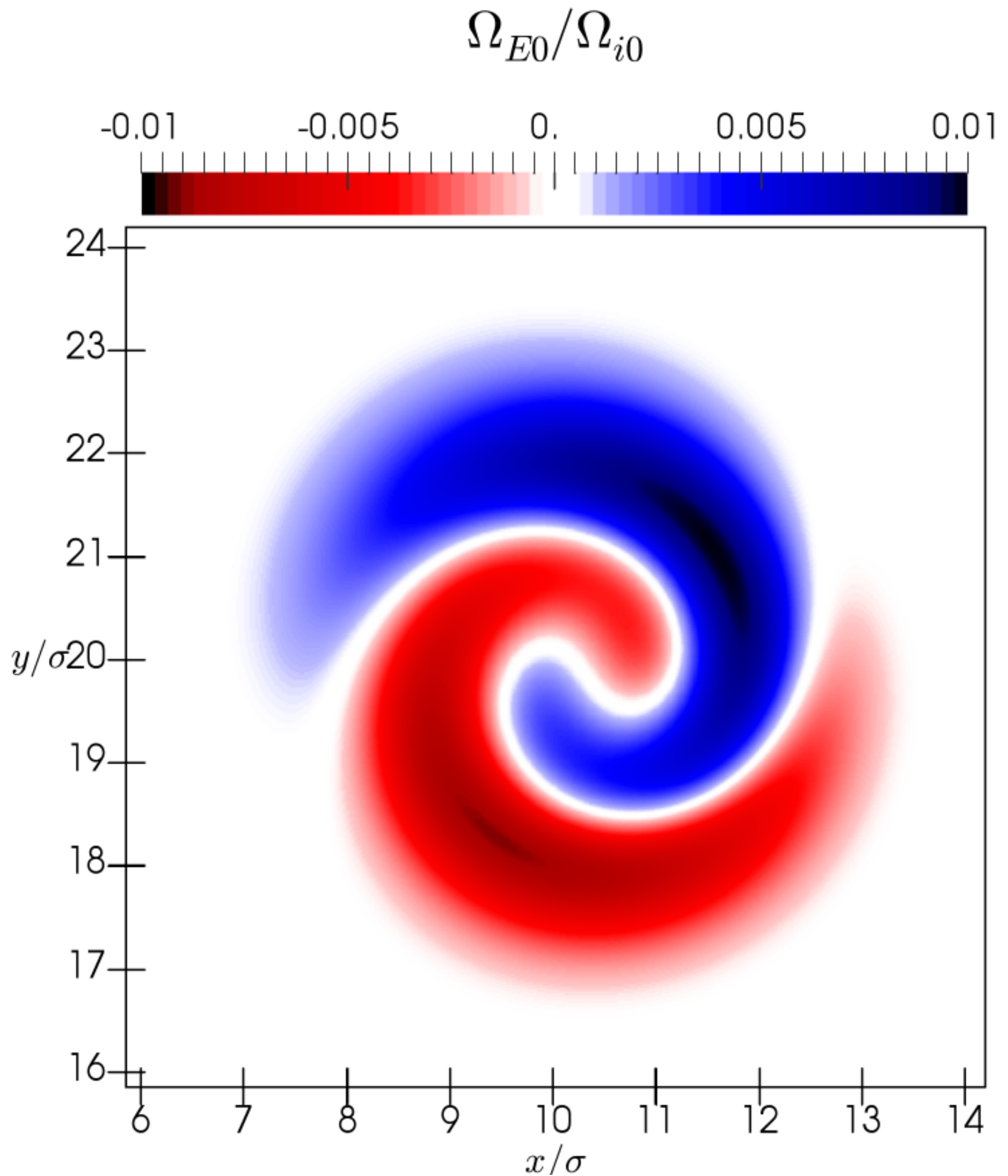}
\caption{The initial rolling up of the $\vec{E}\times \vec{B}$ vorticity $\Omega_{E0}\approx \frac{1}{B_0} \vec{\nabla}_\perp^2 \phi $ is shown for a hot thermal blob with $\tau_i=4$, $A=1$, $\sigma=10$ at two different time steps $t = \left\{16, 
155\right\}\Omega_{i0}^{-1}$.}
\label{fig:vorini}
\end{figure}

Finite ion temperature effects change the behaviour of the blob fundamentally~\cite{madsen11,wiesenberger14}. 
We can see this by comparing the blob structures and positions of~\ref{fig:coldblob} and~\ref{fig:hotblob}. 
FLR corrections arise, which contribute to the polarisation charge density in~\eqref{eq:polarisationeq} 
and advect electrons and ions with different $\vec{E}\times\vec{B}$ drifts $\vec{u}_E$ and $\vec{U}_E$ (cf.~\eqref{eq:uE} and ~\eqref{eq:UE}).
Moreover FLR effects manifest themselves by additional drifts (cf.~\eqref{eq:UE2} and~\eqref{eq:Ueta}) in the gyrofluid moment equations and by the ion diamagnetic vorticity $\Omega_d$ in~\eqref{eq:omegastar}. 
In the initial phase a tilted $\vec{E}\times\vec{B}$-vorticity dipole is generated by a finite ion diamagnetic vorticity $\Omega_d$ and the term  $-\frac{1}{B} \left[\vec{\nabla}_\perp\phi,\vec{\nabla}_\perp \left( \rho_i^2 n_e \right) \right]_{x y}$ in 
the generalised vorticity density~\eqref{eq:voreq} (cf.~\ref{fig:vorini} and~\cite{madsen11}).
Hence, the motion of the blobs is no more purely radial and an additional poloidal motion in the $\vec{\hat{b}} \times \vec{\nabla} B$ direction is observed. 
Subsequently, the tilted $\vec{E}\times\vec{B}$-vorticity dipole rolls itself up and FLR effects lead to the aligning of $\vec{E}\times\vec{B}$ vorticity with ion pressure $p_{i}$. 
The $\vec{E}\times\vec{B}$ shear flow produces a differential rotation, which is 
strongest at the leading edge of the blob.
This effect helps to retain the initial rotational symmetry of the blob and decreases the separation of small eddies. As a consequence
the blob tends to retain its initial Gaussian structure.
This is shown for electron density $n_e$, ion temperature $t_i$
and the $\vec{E}\times\vec{B}$ vorticity 
of an exemplary hot blob in~\ref{fig:hotblob}. 
The electron temperature fields match to a great extent the electron density fields. 
\subsection{Thermal effects on the propagation of a blob}\label{sec:prop}
In this section we investigate the propagation of isothermal and thermal blobs for various ion temperatures, cross-field sizes and initial amplitudes. 
The position of the blob is tracked either by its centre of mass (COM) or by its maximal electron density amplitude.
We define the COM of a blob by
\begin{align}
  \vec{R}_c \equiv \frac{1}{M} \int d\vec{x} \left(n_e-n_{e0}\right)\vec{x}, 
  \end{align}
  \begin{align}
   M \equiv \int d\vec{x} \left(n_e-n_{e0}\right).
\end{align}
The COM displacement $\Delta \vec{R}_c(t)\equiv \vec{R}_c(t)- \vec{R}_c(0)= \left(\Delta X_c(t),\Delta Y_c(t)\right)^T$ is associated with the time integrated perpendicular particle transport 
$\vec{\mathcal{T}}_{n_e} (t) \equiv  \int_0^t dt'\vec{\Gamma}_{n_e} (t')$ via 
\begin{align}
  \vec{\mathcal{T}}_{n_e} (t) \approx   M \Delta \vec{R}_c (t).
\end{align}
Here, we introduced the perpendicular particle transport $ \vec{\Gamma}_{n_e} (t) \equiv   \int d \vec{x} n_e (\vec{u}_E + \vec{u}_{\vec{\nabla} B})$.
The position of the maximal electron density amplitude $\vec{R}_{max}$ is defined implicitly through $ n_e(\vec R_{max}) \geq n_e(\vec x)\ \forall\ \vec x $ and differs from the COM position  $\vec{R}_{c}$ after the initial acceleration phase. 
It is a good measure for the position of the blob front. 
The COM position $\vec{R}_{c}$ and the position of the maximal electron density amplitude $\vec{R}_{max}$ are visualised in~\ref{fig:coldblob} and~\ref{fig:hotblob}. 
The displacement of the maximal electron density amplitude $\vec{R}_{max}$ is defined by $\Delta \vec{R}_{max}(t)\equiv \vec{R}_{max}(t)- \vec{R}_{max}(0)= \left(\Delta X_{max}(t),\Delta Y_{max}(t)\right)^T$.
\subsubsection{Cold ions}\label{sec:coldionsprop}
\begin{figure*}[!ht]
\centering
\begin{subfigure}[!ht]{0.49\textwidth}
\caption{}
    \label{fig:posXvstimescolds10a}
 \includegraphics[trim =0px 0px 0px 0px, clip, scale=0.64]{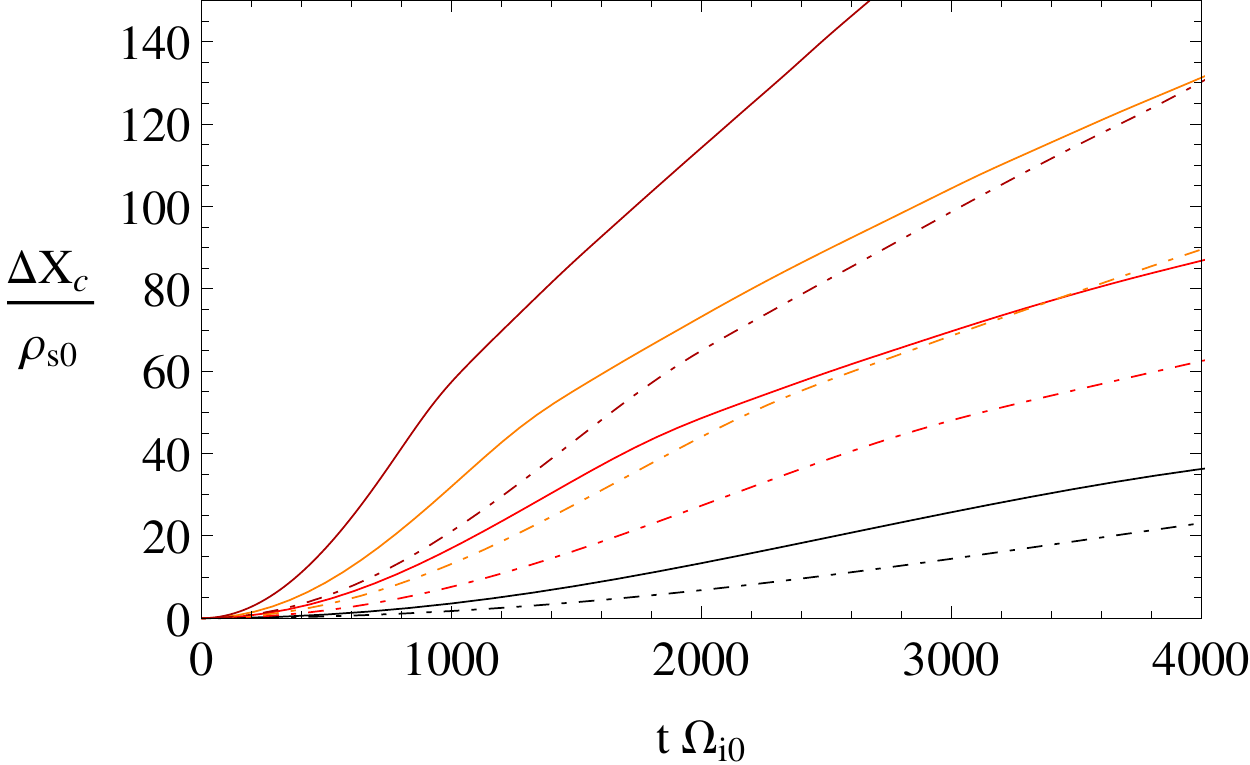}       
\end{subfigure}
\begin{subfigure}[!ht]{0.49\textwidth}
\caption{}
  \label{fig:posXvstimescolds10b}
 \includegraphics[trim = 0px 0px 0px 10px, clip, scale=0.65]{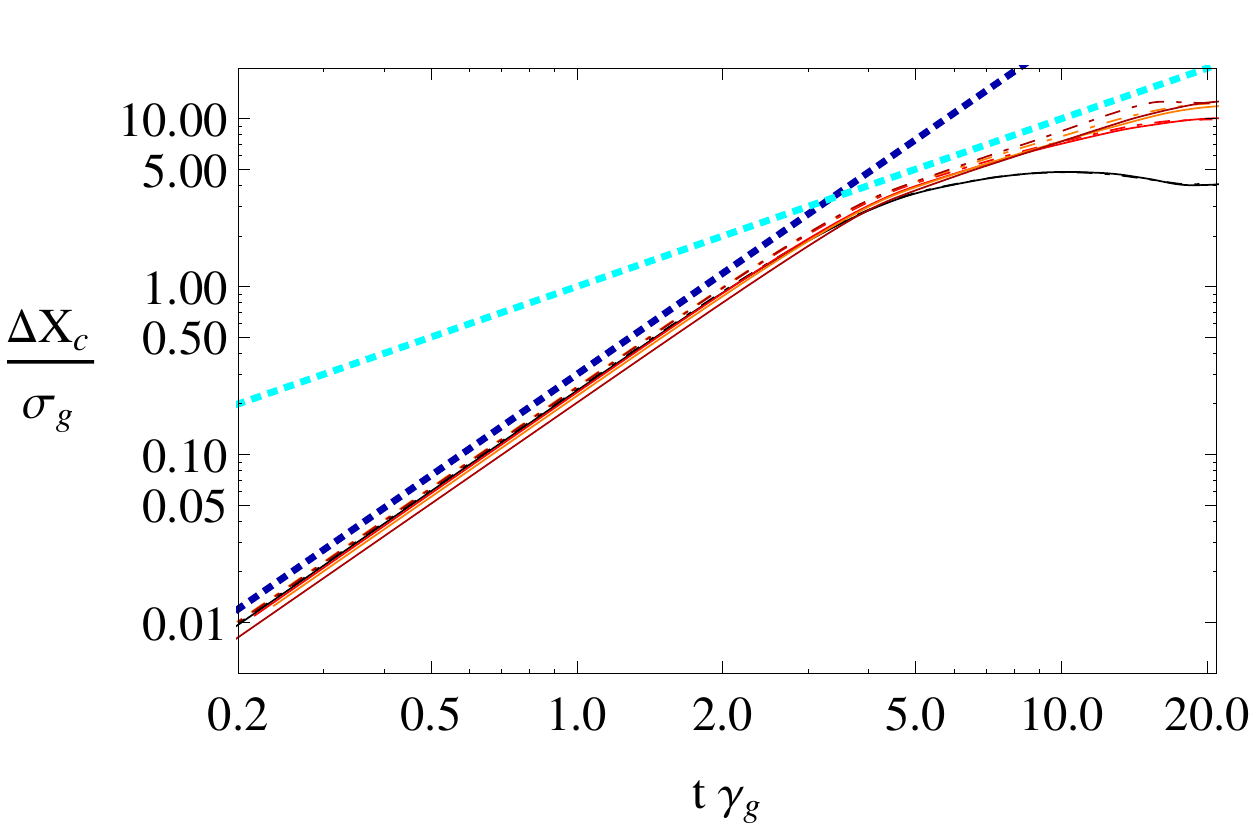}
\end{subfigure}
\caption{The radial displacement of the COM position $\Delta X_c$ as a function of time $t$ is plotted for cold ions $ \tau_i= 0$, cross-field size $\sigma=10$ 
and amplitudes $A = \left\{0.1,0.5,1,2\right\}$ (black, red, orange, dark red). Plot (a) shows that the thermal blobs (solid) travel further into radial direction than isothermal blobs (dot-dashed).
In the normalised double logarithmic plot (b) the dashed dark blue reference line 
$\Delta X_c/\sigma_g \sim t^2 \gamma_g^2$ illustrates the initial constant acceleration phase. The transition to the constant velocity phase 
is represented by the dashed cyan reference line $\Delta X_c/\sigma_g \sim t \gamma_g$.
}
\end{figure*} 
As illustrated in~\ref{sec:blobmotion} the COM motion of cold ion temperature ($\tau_i=0$) blobs is purely radial before the up-down symmetry is broken. 
For $\tau_i=0$ differences in the propagation of isothermal and thermal blobs should be described by our scaling, 
which we used to derive the inertial velocity scaling law of~\eqref{eq:velscalingth}. 
\ref{fig:posXvstimescolds10a} shows that the radial COM displacement $\Delta X_c $ for a given time $t$ is increased by the initial amplitude $A$. 
In order to quantify the amplitude scaling of the radial COM displacement $\Delta X_c$, we 
normalise the radial COM displacement $\Delta X_c$ by the effective blob size $\sigma_g$ and the time by the ideal interchange rate $\gamma_g$.
This must then result in nearly overlapping plots if the velocity scaling estimate of~\eqref{eq:velscalingth} is correct.
This is shown in~\ref{fig:posXvstimescolds10b} and reveals that the radial COM velocities obey 
the velocity scaling estimate of~\eqref{eq:velscalingth} in the cold ion limit. \ref{fig:posXvstimescolds10b} also shows that  
in the initial phase the radial displacement experiences a constant acceleration phase with $\Delta X_c(t) \sim V_\perp \gamma_g t^2 $, 
which is followed by a constant velocity phase with $\Delta X_c(t) \sim V_\perp t $. 
Hence, the time integrated radial particle transport depends in both time phases on the initial electron pressure perturbation $\Delta p_e$.
The initial electron pressure amplitude is proportional to $\Delta p_{e} \sim n_{e0}\Delta t_{e}+t_{e0}\Delta n_e+\Delta n_e \Delta t_{e} $
for a thermal blob whereas it scales like $\Delta p_{e} \sim t_{e0} \Delta n_e$ for an isothermal blob. 
Consequently, thermal blobs travel faster radially than isothermal blobs as depicted by~\ref{fig:posXvstimescolds10a}. 
We discuss the implications of the COM displacement $\Delta \vec{R}_c$ for the (non time integrated) perpendicular particle transport $\vec{\Gamma}_{n_e}$ in~\ref{sec:trans}. 
\subsubsection{Hot ions}
\begin{figure*}[!ht]
\centering
\begin{subfigure}[!ht]{0.49\textwidth}
\caption{}
    \label{fig:posXvstimeshots10a}
    \includegraphics[trim = 0px 0px 0px 0px, clip, scale=0.64]{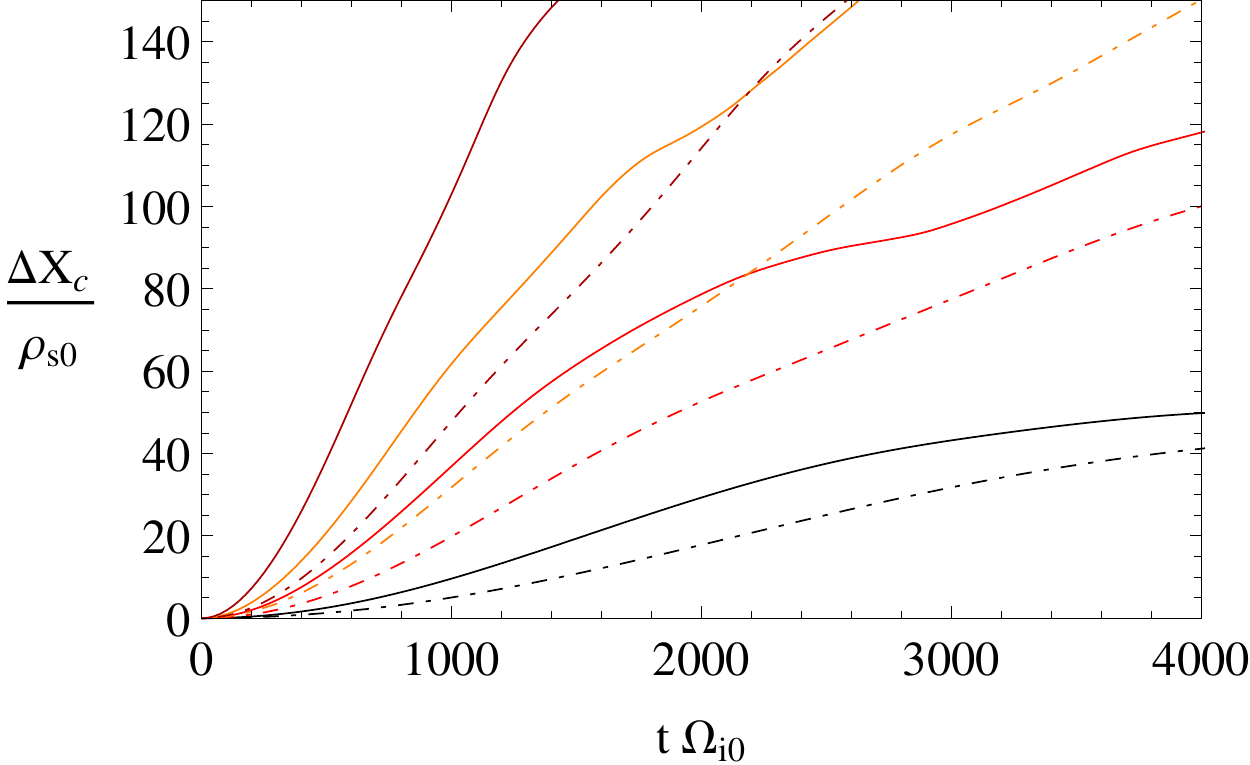}
\end{subfigure}  
\begin{subfigure}[!ht]{0.49\textwidth}
\caption{}
    \label{fig:posXvstimeshots10b}
\includegraphics[trim = 0px 0px 0px 10px, clip, scale=0.65]{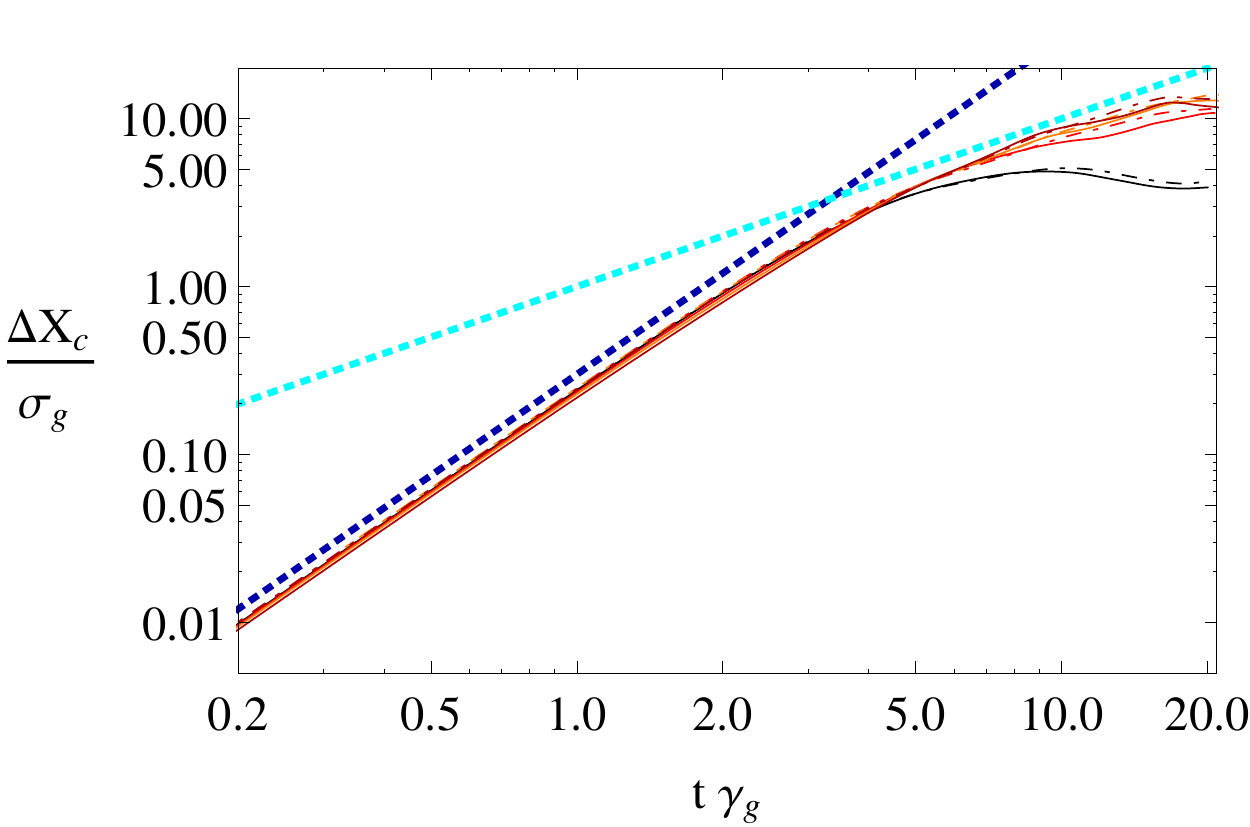}
\end{subfigure}  
\caption{ The radial displacement of the COM position $\Delta X_c$ as a function of time $t$ is plotted for hot ions $ \tau_i= 2$, cross-field size $\sigma=10$
and amplitudes $A = \left\{0.1,0.5,1,2\right\}$ (black, red, orange, dark red). Plot (a) shows that the thermal blobs (solid) travel further into radial direction than isothermal blobs (dot-dashed).
In the normalised double logarithmic plot (b) the dashed dark blue reference line 
$\Delta X_c/\sigma_g \sim t^2 \gamma_g^2$ shows the initial constant acceleration phase. The transition to the constant velocity phase 
is represented by the dashed cyan reference line $\Delta X_c/\sigma_g \sim t \gamma_g$.
}
\label{fig:posXvstimeshots10}
\end{figure*}
\begin{figure}[!ht]
\centering
\hspace{ 0.05\textwidth}isothermal \hspace{0.15\textwidth} thermal \newline
\includegraphics[trim = 0px 0px 0px 0px, clip, scale=0.32]{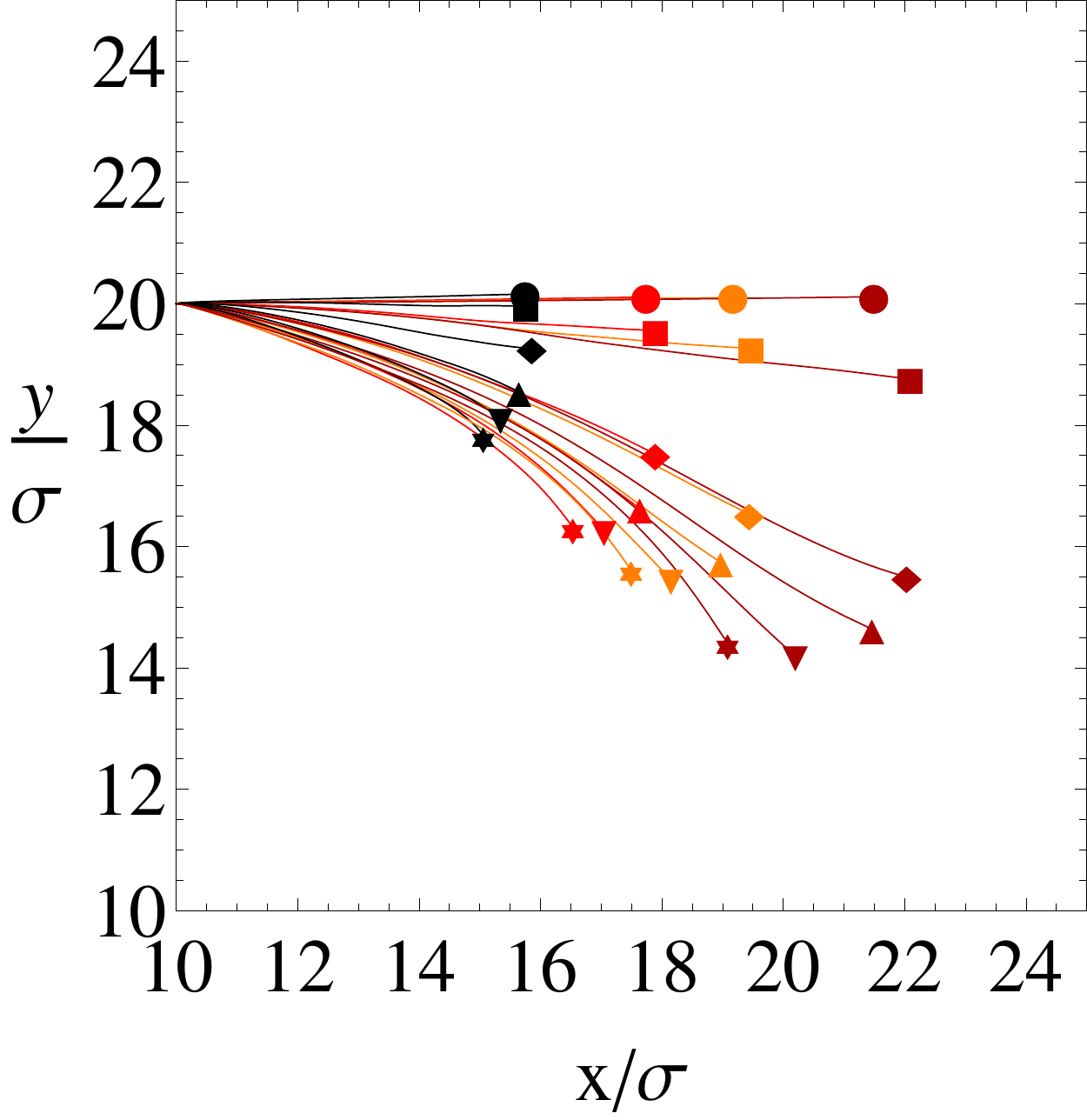}
\includegraphics[trim = 0px 0px 0px 0px, clip, scale=0.32]{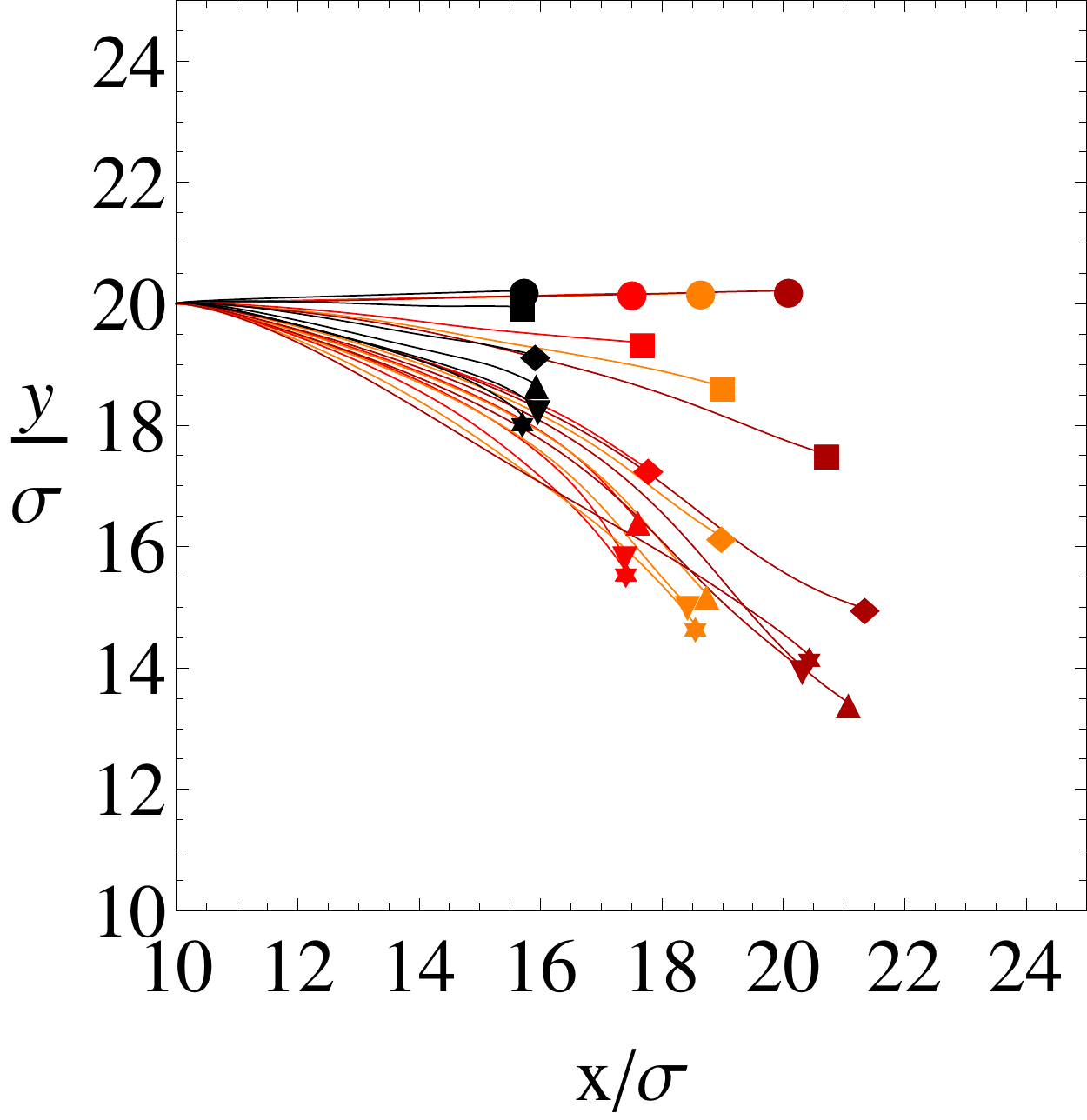}

\includegraphics[trim = 0px 0px 0px 0px, clip, scale=0.32]{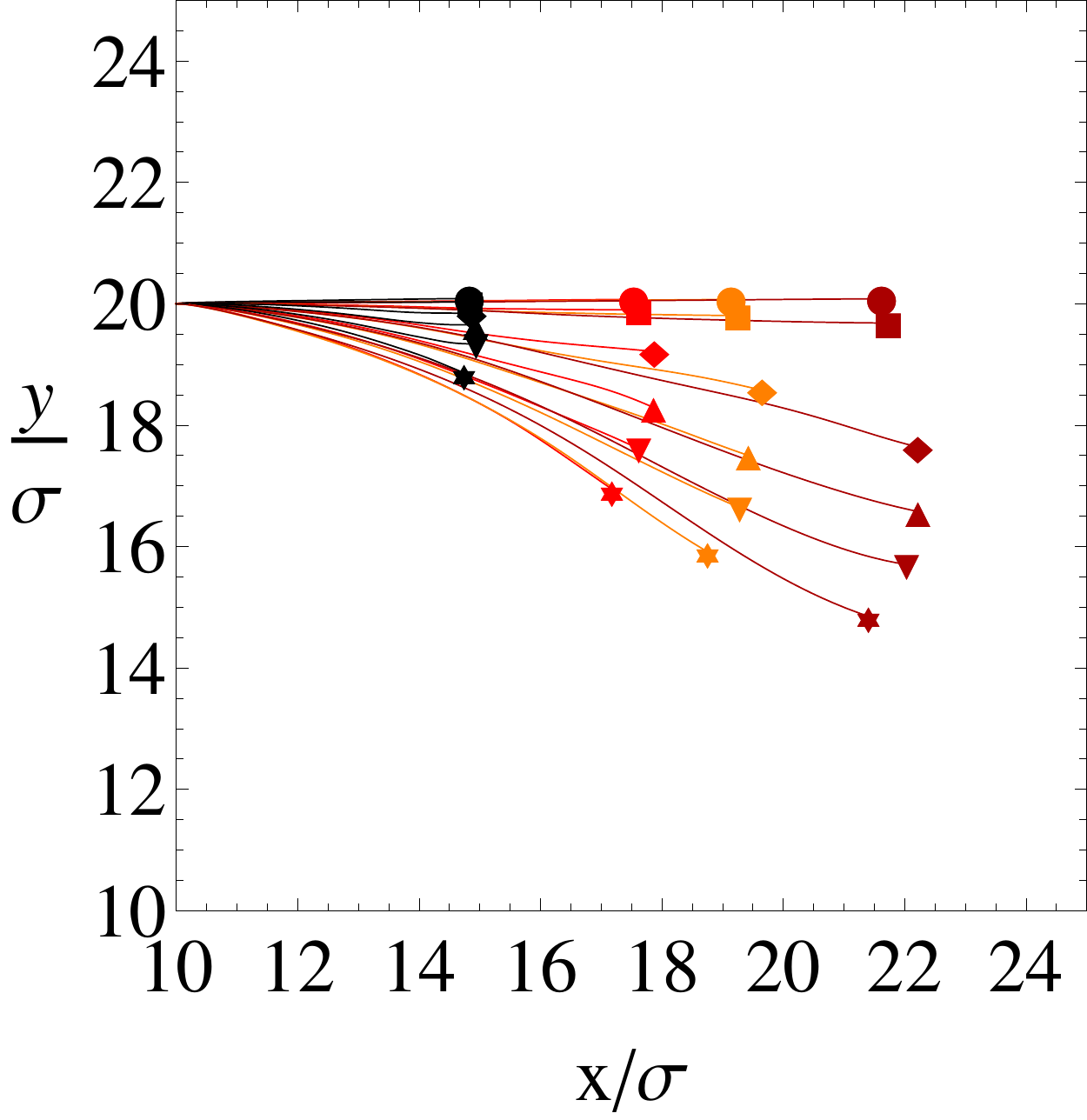}
\includegraphics[trim = 0px 0px 0px 0px, clip, scale=0.32]{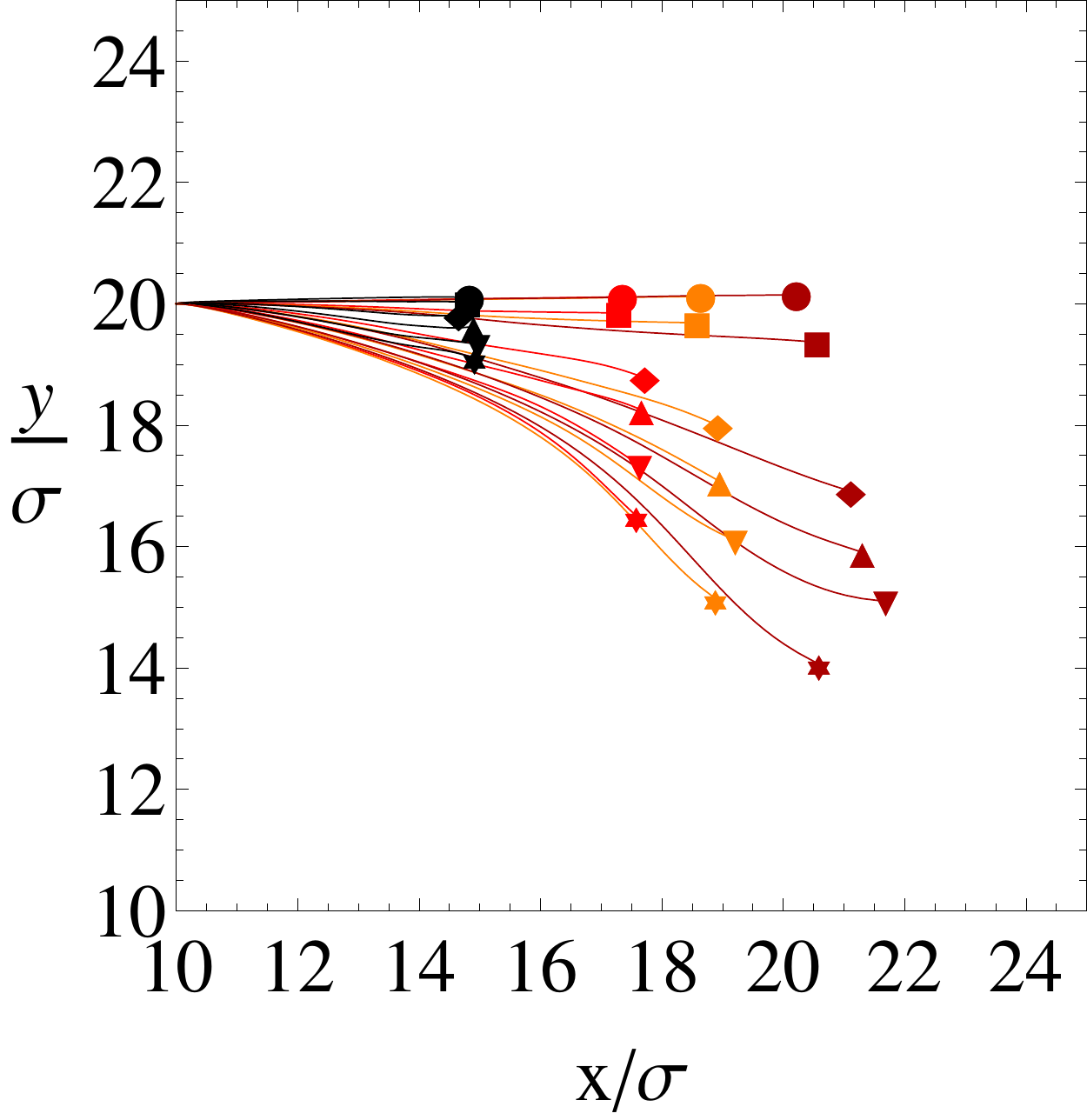}

\includegraphics[trim = 0px 0px 0px 0px, clip, scale=0.32]{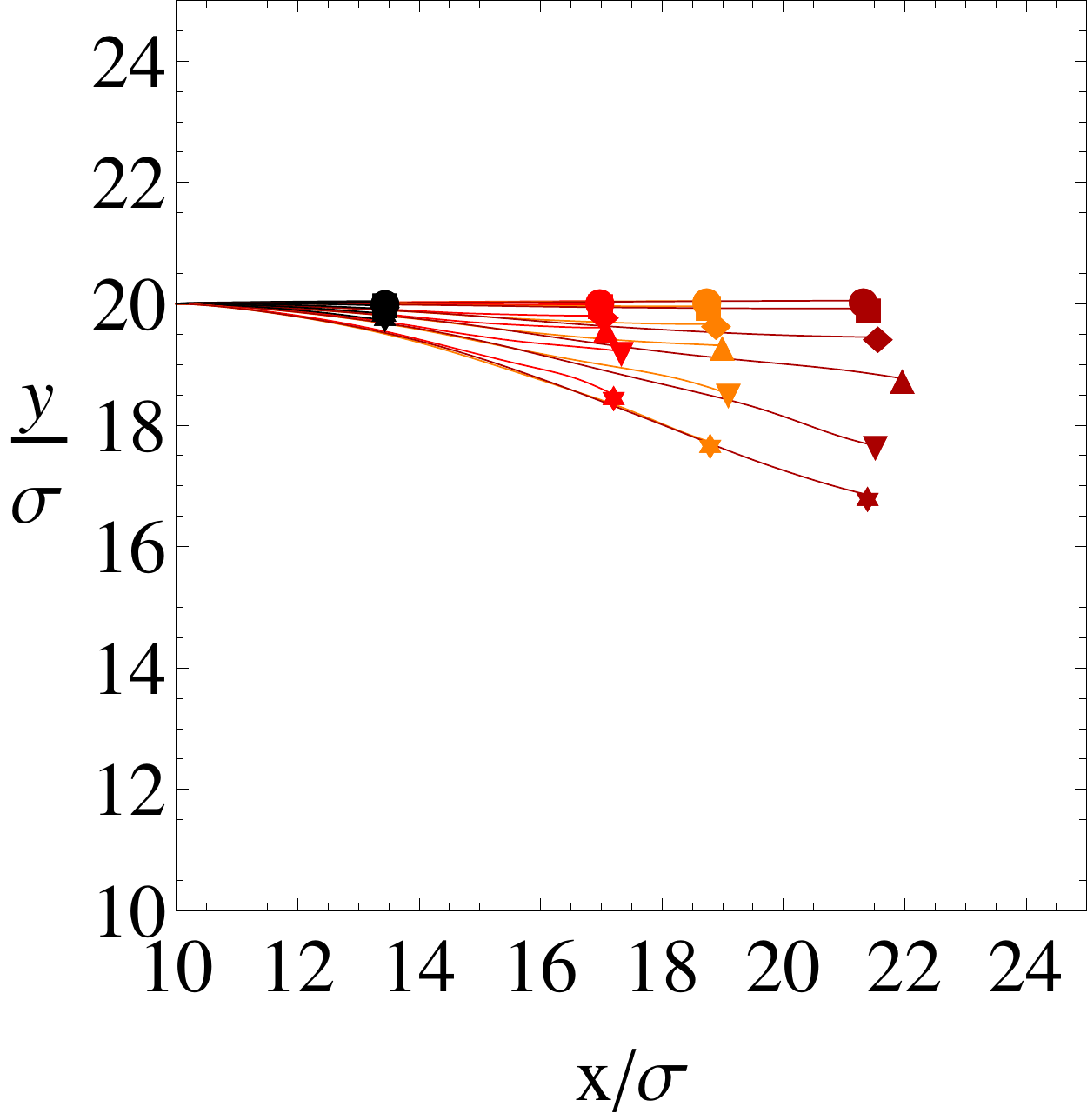}
\includegraphics[trim = 0px 0px 0px 0px, clip, scale=0.32]{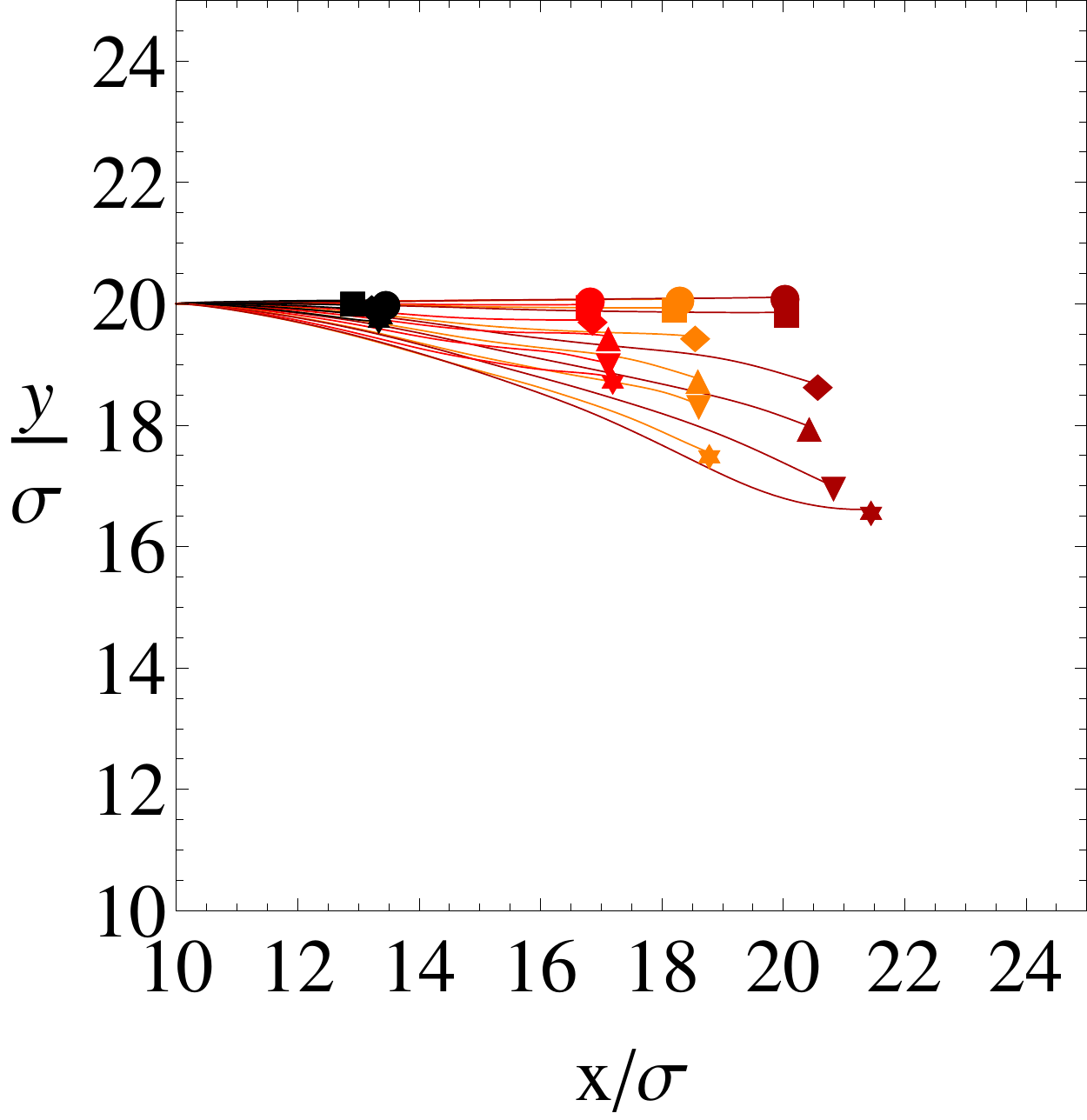}
\caption{The trajectory of the COM position $\vec{R}_c$  at $t=8 \gamma_g^{-1}$ 
for $ \tau_i= \left\{0,0.1,0.5,1,2,4\right\}$ (circle, square, diamond, up triangle, down triangle, star) 
and $A = \left\{0.1,0.5,1,2\right\}$ (black, red, orange, dark red) in case of isothermal (left) and thermal (right) blobs  is pictured. 
The blob cross-field size is $\sigma=5$ (first row), $\sigma=10$ (second row) and $\sigma=20$ (third row). 
The poloidal displacement $\Delta Y_c$ of thermal blobs is enhanced for finite ion background temperature and high amplitudes in comparison to isothermal blobs.}
\label{fig:maxtracki1}
\end{figure}

Finite ion temperature effects ($\tau_i>0$) redistribute the total COM momentum into the radial and poloidal directions. 
An ion temperature perturbation $\Delta t_{i}$ enhances the total (radial and poloidal) COM displacement $|\Delta \vec{R}_c|\equiv \sqrt{\Delta X_c^2 + \Delta Y_c^2 }$ compared to isothermal blobs. 
This behaviour is shown for the radial COM displacement $\Delta X_c$ in~\ref{fig:posXvstimeshots10a}. 
The normalisation by the effective cross-field size $\sigma_g$ and the ideal interchange rate $\gamma_g$ is depicted in~\ref{fig:posXvstimeshots10b}.  
This figure shows that the normalised radial COM displacement $\Delta X_c/\sigma_g$ of isothermal and thermal blobs nearly coincide in the constant acceleration and velocity phases. 
Hence, the radial COM velocities $V_{c,x}$ are captured by the inertial velocity scaling estimate $V_\perp$ of~\eqref{eq:velscalingth}.
Since this estimate is only suitable for the radial COM velocities, modifications for total and poloidal
velocity estimates are required. In~\ref{sec:trans} we show that the total and poloidal velocity scaling laws can 
be adjusted with the help of the FLR strength parameter $\Theta$.
\\
In the following we track the radial and poloidal COM position $\vec{R}_{c}$ and investigate the blob propagation for the complete parameter space.\\
In~\ref{fig:maxtracki1} we clarify the dependence of the COM displacement $\vec{R}_{c}$ on the main parameters ($\sigma,\tau_i,A$) for isothermal and thermal blobs. 
We observe that the radial and poloidal COM displacement $\Delta X_{c}$ and $\Delta Y_{c}$ increases with amplitude $A$ over the complete parameter space. 
This is in accordance with the scaling estimates of the radial COM displacement $\Delta X_c$, presented in~\ref{sec:coldionsprop}, and holds similarly for the total COM displacement $|\Delta \vec{R}_{c}|$. 
The poloidal COM displacement $\Delta Y_{c}$ is decreased by the cross-field size $\sigma$ and is also enhanced by the ratio of ion to electron background temperature $\tau_i$. 
In addition, a finite ion temperature perturbation $\Delta t_{i}$ enhances the poloidal COM displacement $\Delta Y_{c}$ especially for high amplitudes $A$ in comparison to isothermal blobs with constant ion temperature $t_{i0}$.
This is in line with the FLR effect strength parameter $\Theta$ of~\eqref{eq:flrstrengthth} and is further discussed in~\ref{sec:trans}.
We ascribe the small increase of poloidal motion to dynamic FLR effects. 
We do so because a finite ion temperature perturbation 
$\Delta t_i$ together with dynamic FLR effects $\rho_i(T_i, B )$ result in the FLR effect strength parameter 
$\Theta\sim\sqrt{\Delta p_{i}^2 /\left((\Delta p_{e}+\Delta p_{i})  p_{e0}\right)}$.
This exceeds its isothermal, constant FLR effect limit equivalent 
$\Theta\sim \sqrt{t_{i0}^2\Delta n_{e} /\left((t_{e0}+ t_{i0})  p_{e0}\right)}$ for all studied parameters. 
This difference stems from either constant or dynamic FLR effects in the polarisation~\eqref{eq:polarisationeq}.
In case of ion temperature variations $\Delta t_{i}$ dynamic FLR effects appear in the polarisation~\eqref{eq:polarisationeq} via the gyroaveraging operator $\Gamma_{i,1}^\dagger$
whereas in the isothermal limit constant FLR effects enter the polarisation~\eqref{eq:polarisationeq} via the gyroaveraging operator $\Gamma_{i,1}$. 
In order to ensure that the poloidal motion depends on constant or dynamic FLR effects, 
we repeated the simulations with temperature dynamics $(t_e,T_i) \neq (t_{e0}, T_{i0})$ 
but constant FLR effects $\rho_i(T_{i0},B_0)$.
The related numerical results and the influence of FLR effects on the poloidal particle transport of a blob is presented in~\ref{sec:trans}.
\subsection{Thermal effects on blob compactness}\label{sec:comp}
\begin{figure}[!ht]
\centering
\includegraphics[trim = 0px 0px 0px 0px, clip, scale=0.64]{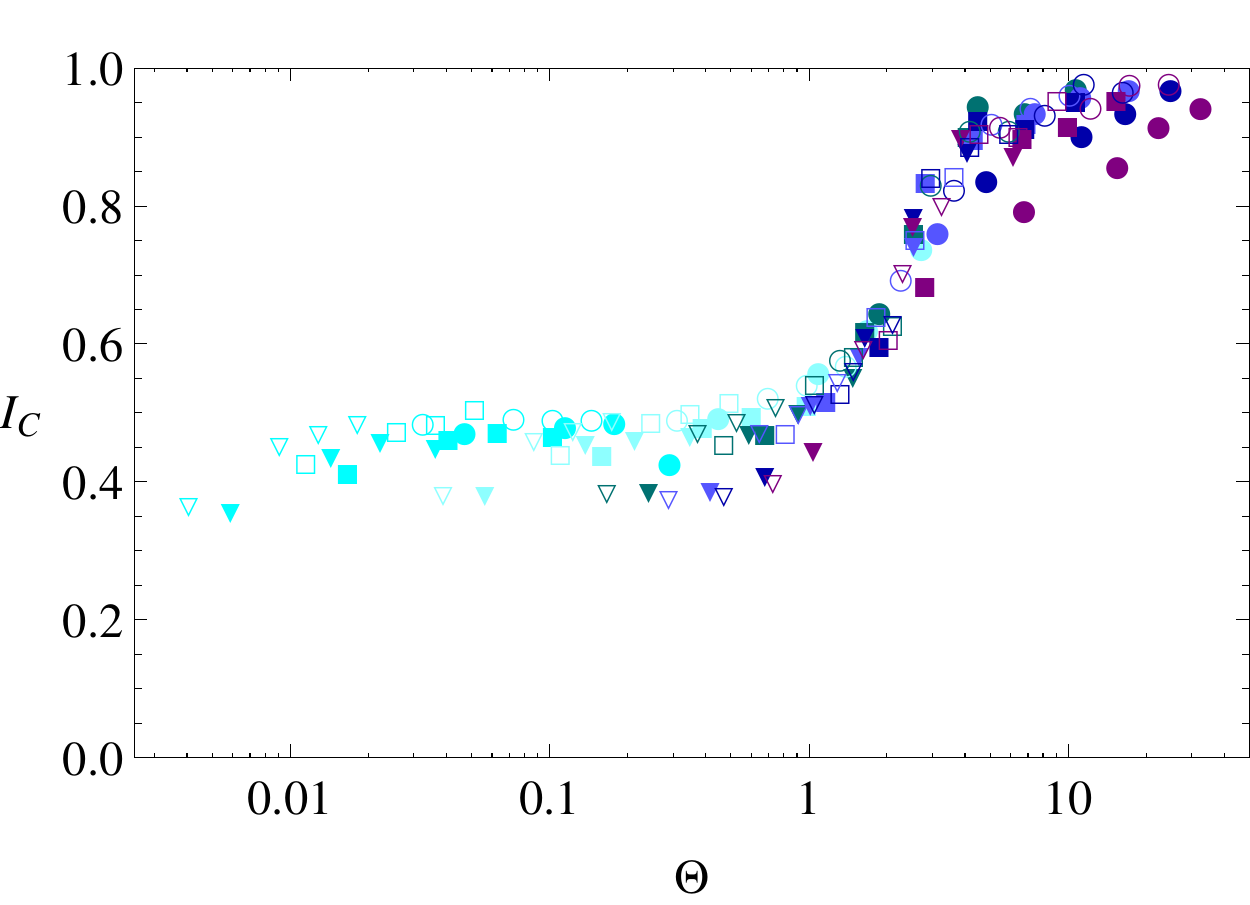}
\caption{The compactness of electron density at the time $t=4\gamma_g^{-1}$ as a function of the FLR effect strength is depicted in a logarithmic plot. 
The symbols indicate the blob size $\sigma = \left\{5,10,20 \right\} $ (circle, square, down triangle) and the colour the ratio of ion to electron background 
temperature $\tau_i =  \left\{0,0.1,0.5,1,2,4 \right\} $ 
(cyan, light cyan, dark cyan, light blue, dark blue, purple). 
Filled symbols are thermal blobs and empty symbols are isothermal blobs.}
\label{fig:compall}
\end{figure}

\begin{figure}[!ht]
\centering
\includegraphics[trim = 0px 0px 0px 0px, clip, scale=0.65]{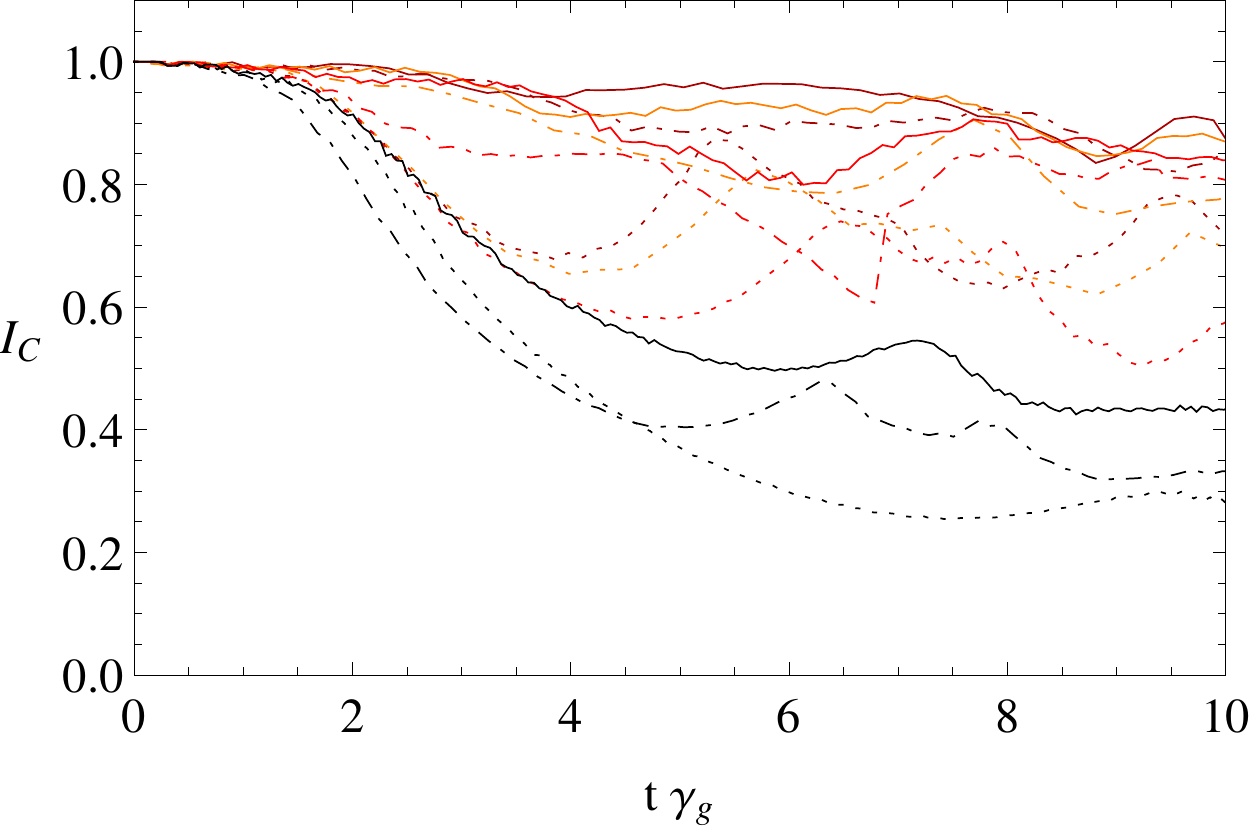}
\caption{
The compactness $I_C$ as a function of normalised time $t\gamma_g$ is shown for hot ions $\tau_i=2$, cross-field size $\sigma=10$ and amplitudes $A=\left\{0.1,0.5,1,2\right\}$ (black, red,orange,dark red). 
Thermal blobs with dynamic FLR effects (solid) remain more compact than thermal or isothermal blobs with constant FLR effects (dotted, dot-dashed). }
\label{fig:compactnessFLR}
\end{figure}

The ability of a blob to retain its initial shape is given by the blob compactness $ I_C(t)$ \cite{madsen11}
\begin{align}\label{eq:compactness}
 I_C(t) \equiv \frac{\int d\vec{x} \left(n_e (\vec{x},t)-n_{e0}\right) h(\vec{x},t)}{\int d\vec{x} \left(n_e (\vec{x},0)-n_{e0}\right)h(\vec{x},0)},
\end{align}
with the Heaviside function 
\begin{align}\label{eq:heavi}
h(\vec{x},t) =\begin{cases}
1,&\text{for} \hspace{3 mm} \left\|\vec{x}-\vec{R}_{max}(t)\right\|^2  <\sigma^2\\
0,&\text{else}\\
\end{cases}
\end{align}
and with $\vec{R}_{max}(t)$ defined as the position of the maximum electron density amplitude.
The lowering of compactness is accompanied by the loss of mass via collisional dissipation and turbulent mixing and stretching
whereas a finite Larmor radius contributes to long lived coherent compact structures.
Consequently, from~\ref{fig:coldblob} and \ref{fig:hotblob} we expect low compactness for cold blobs and high compactness for hot blobs. \\
The maximal radial particle transport of a blob is determined by the blob compactness and the maximal radial velocity. 
As a result we plot the blob compactness at the time $t=4\gamma_g^{-1}$, which approximately coincides with the first maximum of the radial velocity, versus 
the FLR effect strength  $\Theta$ (cf.~\eqref{eq:flrstrengthth}) in~\ref{fig:compall}. 
Note that for high $\tau_i$ the maximal radial velocity is often reached at the second peak in time (see e.g.~\ref{fig:velvstimeshots10}), 
which is accompanied by a small drop in compactness $I_C$.
In~\ref{fig:compall} a smooth transition to highly compact blobs is obtained at $\Theta \approx1$. 
Above this threshold typically only 10 percent of the initial particle density is lost whereas below the threshold up to 50 percent of the initial particle
density is dissipated. 
For constant $\tau_i$ the transition to the ion diamagnetic vorticity dominated regime  ($\Theta \gg 1$) is reached by decreasing the cross-field size $\sigma$ or increasing the amplitude of the blob.
The thermal and isothermal blobs approach similar values of compactness $I_C$ over the complete \(\Theta\) range. 
We note here that for the sake of clarity the zero ion temperature value $\tau_i=0$ was replaced by $\tau_i=0.01$ in order to compute a finite FLR strength $\Theta$.
\\
On the other hand, the FLR strength parameter $\Theta$ of thermal blobs is larger than of isothermal blobs for constant cross-field size $\sigma$, amplitude $A$ and ion to electron background ratio $\tau_i$. 
As an example for $\sigma=10$, $A=0.5$ and $\tau_i=2$ the FLR strength parameter is $\Theta\approx 5$ for thermal blobs and $\Theta\approx 3$ for isothermal blobs. 
Consequently,~\ref{fig:compall} shows that thermal blobs remain more compact than isothermal blobs. 
This is underlined by~\ref{fig:compactnessFLR}, which reveals that thermal blobs with dynamic FLR effects lose less mass than thermal or isothermal blobs with constant FLR effects.
This is due to differences in the polarisation charge density for constant and dynamic FLR effects, which is discussed in~\ref{sec:polarisation}. 
Since the polarisation charge density is associated to the $\vec{E}\times \vec{B}$ vorticity, dynamic FLR effects 
lead to stronger $\vec{E}\times \vec{B}$ shear flows in the blob edge in comparison to constant FLR effects. These $\vec{E}\times \vec{B}$ shear flows help to retain the initial blob shape (cf.~\ref{sec:blobmotion}). 
As a result the compactness of blobs with a dynamic gyro-radius is higher than of blobs with a constant gyro-radius. 
\subsection{Thermal effects on the particle transport of a blob}\label{sec:trans}
In the following we use the COM velocity $\vec{V}_c$ as a measure for the perpendicular particle transport 
\begin{align}
  \vec{\Gamma}_{n_e} (t)  \approx M \vec{V}_c.
\end{align}
The COM velocity is given by:
\begin{align}
  \vec{V}_c \equiv \frac{d \vec{R}_c}{d t}.
\end{align}
In the remainder of this section the influence of the ratio of ion to electron background temperature $\tau_i$, cross-field size $\sigma$ and 
initial amplitude $A$ on the transport of isothermal and thermal blobs is studied. 
\subsubsection{Cold ions}
\begin{figure}[!ht]
\centering
\includegraphics[trim = 0px 0px 0px 0px, clip, scale=0.64]{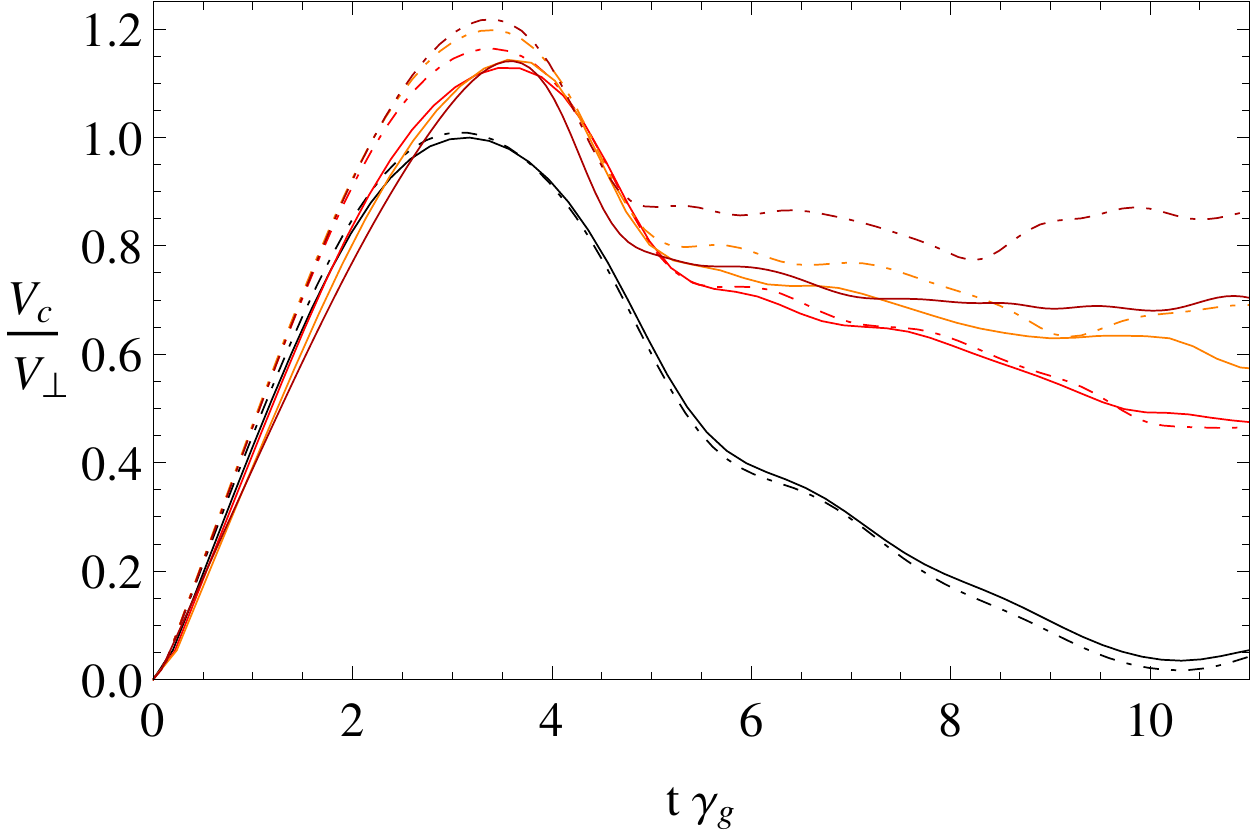}
\caption{ 
The normalised total COM velocity $V_c/V_\perp$ as a function of normalised time $t \gamma_g$ is plotted for cold ions $ \tau_i= 0$, cross-field size $\sigma=10$
and amplitudes $A = \left\{0.1,0.5,1,2\right\}$ (black, red, orange, dark red). 
The theoretical maximal velocity $V_\perp$ is fitted within twenty percent and the global ideal interchange rate $\gamma_g$ predicts the time period in which the COM velocity reaches its maximum up to a constant factor.
}
\label{fig:velvstimescolds10}
\end{figure}
\begin{figure}[!ht]
\centering
\includegraphics[trim = 0px 0px 0px 0px, clip, scale=0.64]{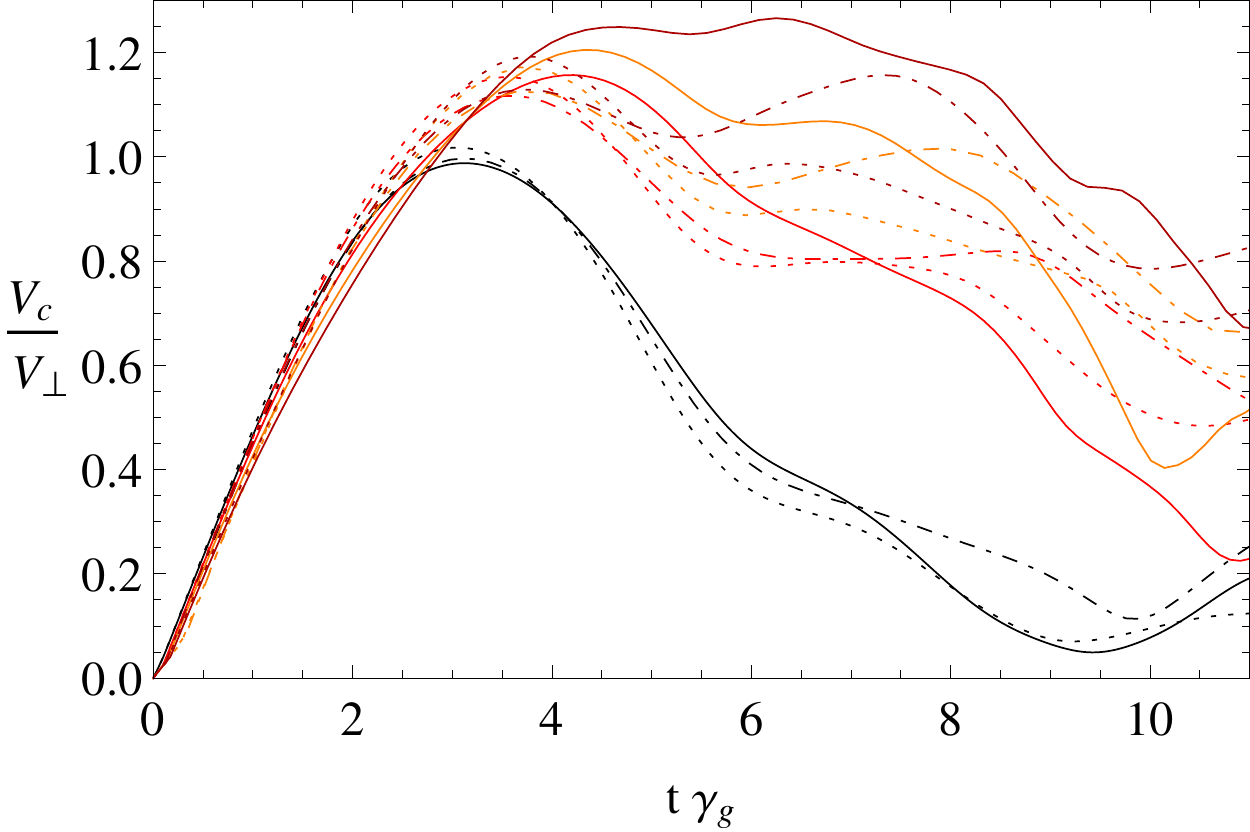}
\caption{
The normalised total COM velocity  $V_c/V_\perp$ as a function of normalised time $t \gamma_g$ is plotted for hot ions $ \tau_i= 2$, cross-field size $\sigma=10$
and amplitudes $A = \left\{0.1,0.5,1,2\right\}$ (black, red, orange, dark red). 
Thermal blobs (solid and dotted) and isothermal blobs (dot-dashed) show up similar behaviour of the total COM velocity. 
}
\label{fig:velvstimeshots10c}
\end{figure}

In case of zero ion temperature the radial and total transport coincide. In~\ref{fig:velvstimescolds10} the maximal COM velocity is reached after an 
initial constant acceleration phase and is captured within twenty percent by the inertial velocity scaling law $V_\perp$ of~\eqref{eq:velscalingth}. The time period to reach the
maximal COM velocity scales with the global ideal interchange rate $\gamma_g$ of~\eqref{eq:gammascalingthg}.
\begin{figure*}[!ht]
\centering
\begin{subfigure}[!ht]{0.49\textwidth}
\caption{}
\label{fig:velvstimeshots10a}
\includegraphics[trim = 0px 0px 0px 0px, clip, scale=0.66]{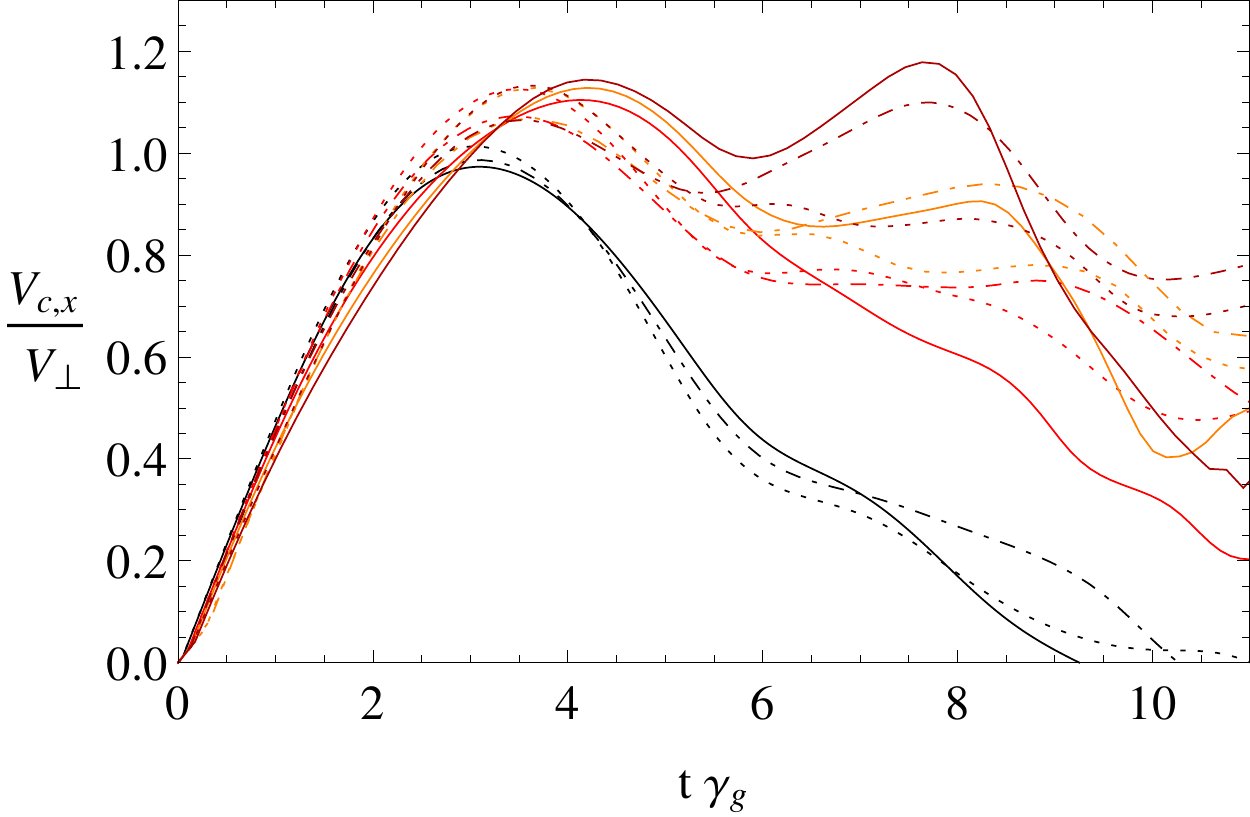}
\end{subfigure}
\begin{subfigure}[!ht]{0.49\textwidth}
\caption{}
\label{fig:velvstimeshots10b}
\includegraphics[trim = 0px 0px 0px 0px, clip, scale=0.66]{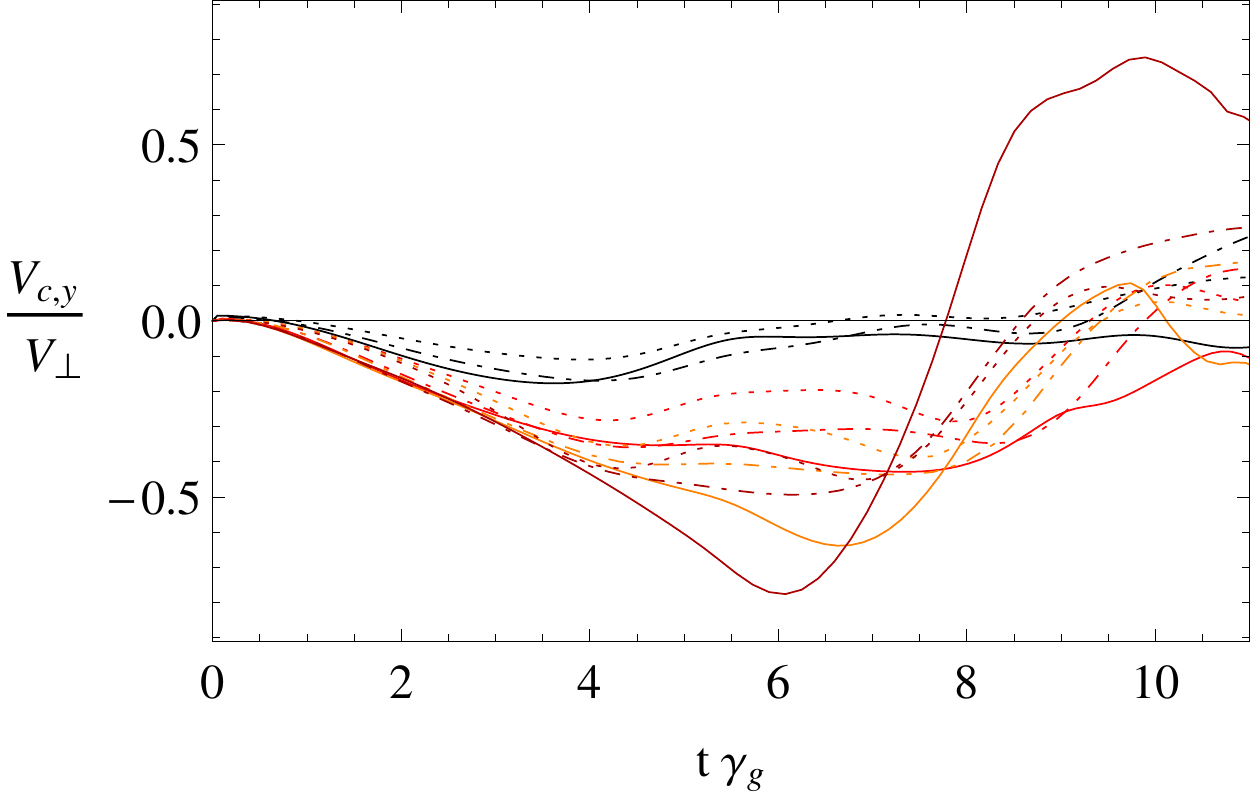}
\end{subfigure}
\caption{
The normalised radial (a) and poloidal (b) COM velocity $V_{c,x}/V_\perp$ and $V_{c,y}/V_\perp$ as a function of normalised time $t \gamma_g$  is plotted for hot ions $ \tau_i= 2$, cross-field size $\sigma=10$
and amplitudes $A = \left\{0.1,0.5,1,2\right\}$ (black, red, orange, dark red). (a)
Thermal blobs with dynamics FLR effects (solid) and thermal blobs with constant FLR effects (dotted) feature similar normalised radial COM velocities as isothermal blobs (dot-dashed). (b) 
The normalised minimal poloidal COM velocities of thermal blobs with dynamic FLR effects (solid) are lower than those of blobs with constant FLR effects (dotted and dot-dashed).
}
\label{fig:velvstimeshots10}
\end{figure*}
\begin{figure*}[!ht]
\centering
\hspace{0.12\textwidth}$\sigma=5$ \hspace{0.28\textwidth}$\sigma=10$  \hspace{0.26\textwidth}$\sigma=20$ \newline
\includegraphics[trim = 0px 0px 0px 0px, clip, scale=0.44]{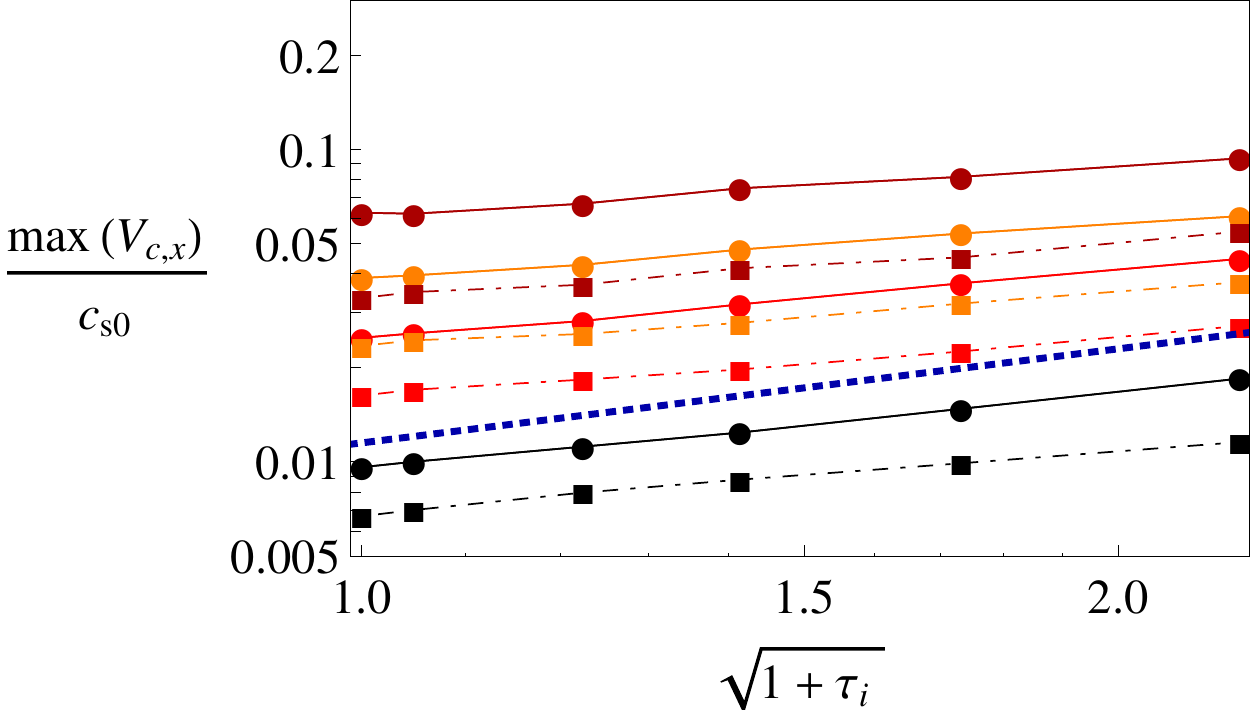} 
\includegraphics[trim = 0px 0px 0px 0px, clip, scale=0.44]{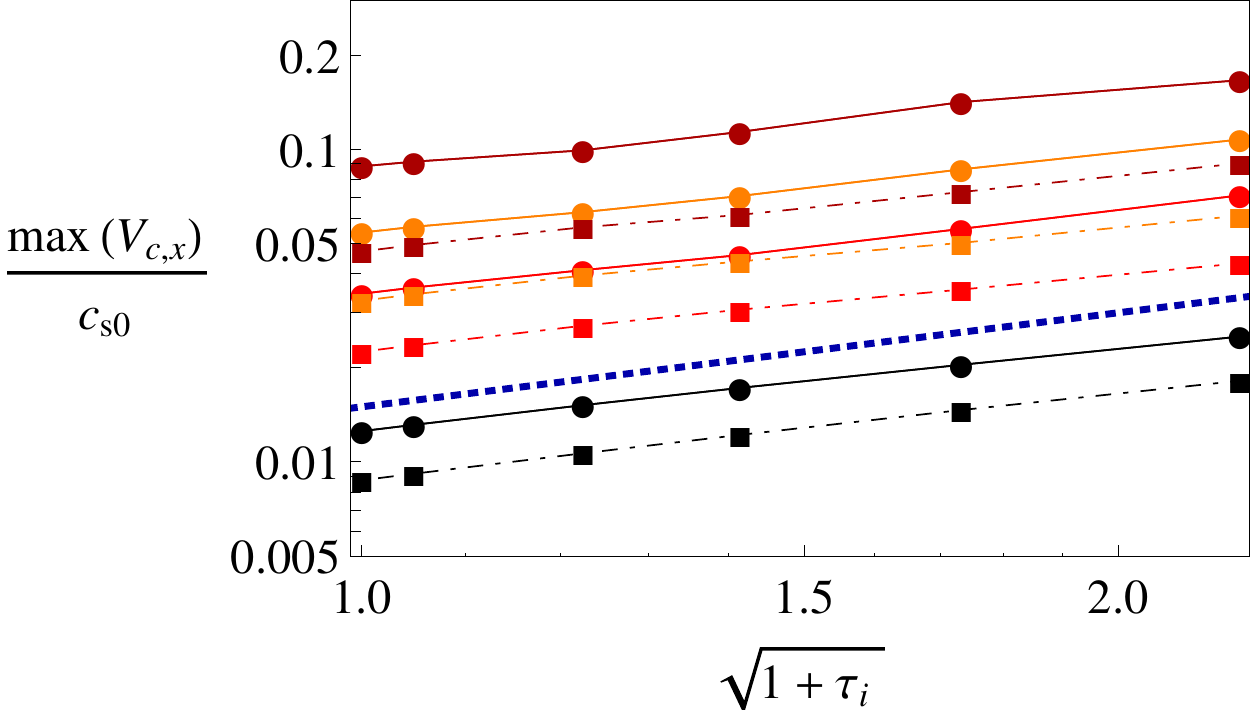} 
\includegraphics[trim = 0px 0px 0px 0px, clip, scale=0.44]{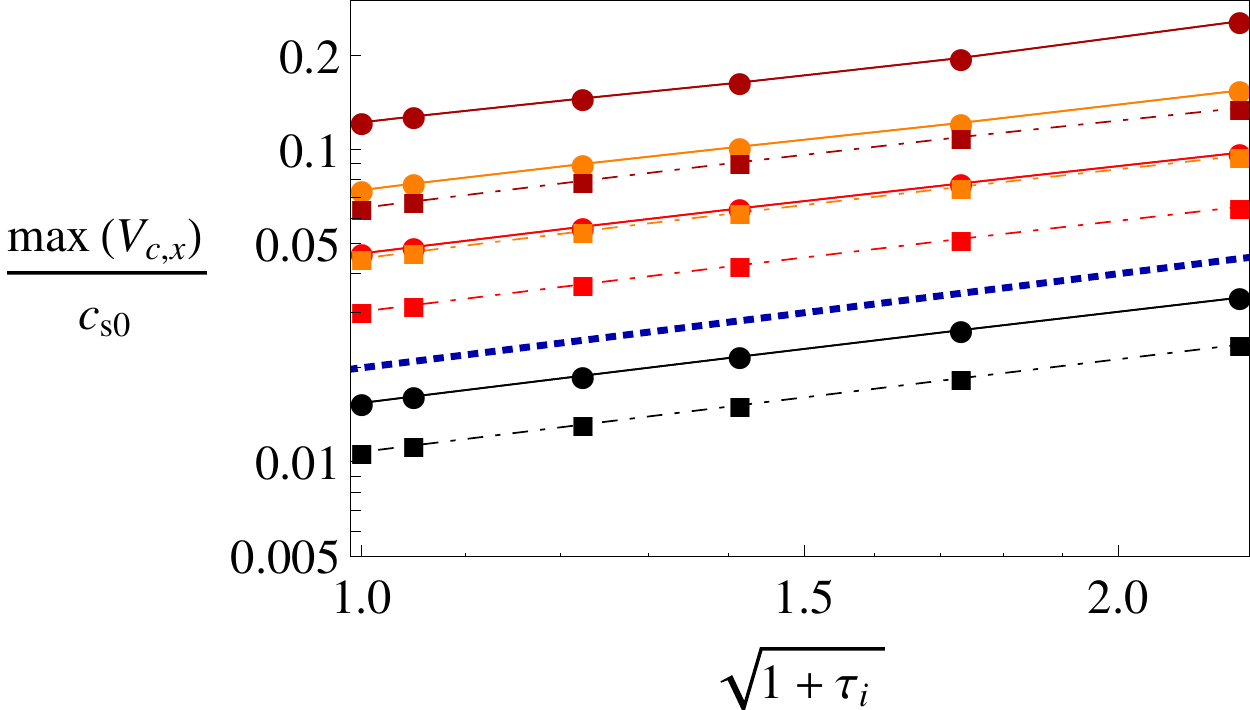}
\caption{The ion to electron background temperature ratio $\tau_i$ dependence of the maximal radial COM velocity $\textrm{max}(V_{c,x})$ for four different amplitudes $A = \left\{0.1,0.5,1,2\right\}$ (black, red, orange, dark red) and three different cross-field sizes
$\sigma$ are shown. The solid lines represent thermal blobs and the dot-dashed lines isothermal blobs. The dashed dark blue line represents the $V_\perp/c_{s0}\sim \sqrt{1+\tau_i}$ reference line.}
\label{fig:vmaxvstau}
\end{figure*}
\\
In~\ref{fig:velvstimeshots10c} the time dependence of the normalised total COM velocities $V_c/V_\perp$ are plotted. The maximal total COM velocities $V_c$
agree with the velocity scaling estimates of~\eqref{eq:velscalingth} but are slightly underestimated by the velocity scaling estimates of~\eqref{eq:velscalingth} in case of dynamic FLR effects.
The time period to reach this maximum is enhanced for high blob amplitudes $A$. 
\begin{figure*}[!ht]
\centering
\hspace{0.12\textwidth}$\sigma=5$ \hspace{0.28\textwidth}$\sigma=10$  \hspace{0.26\textwidth}$\sigma=20$ \newline
\includegraphics[trim = 0px 0px 0px 0px, clip, scale=0.44]{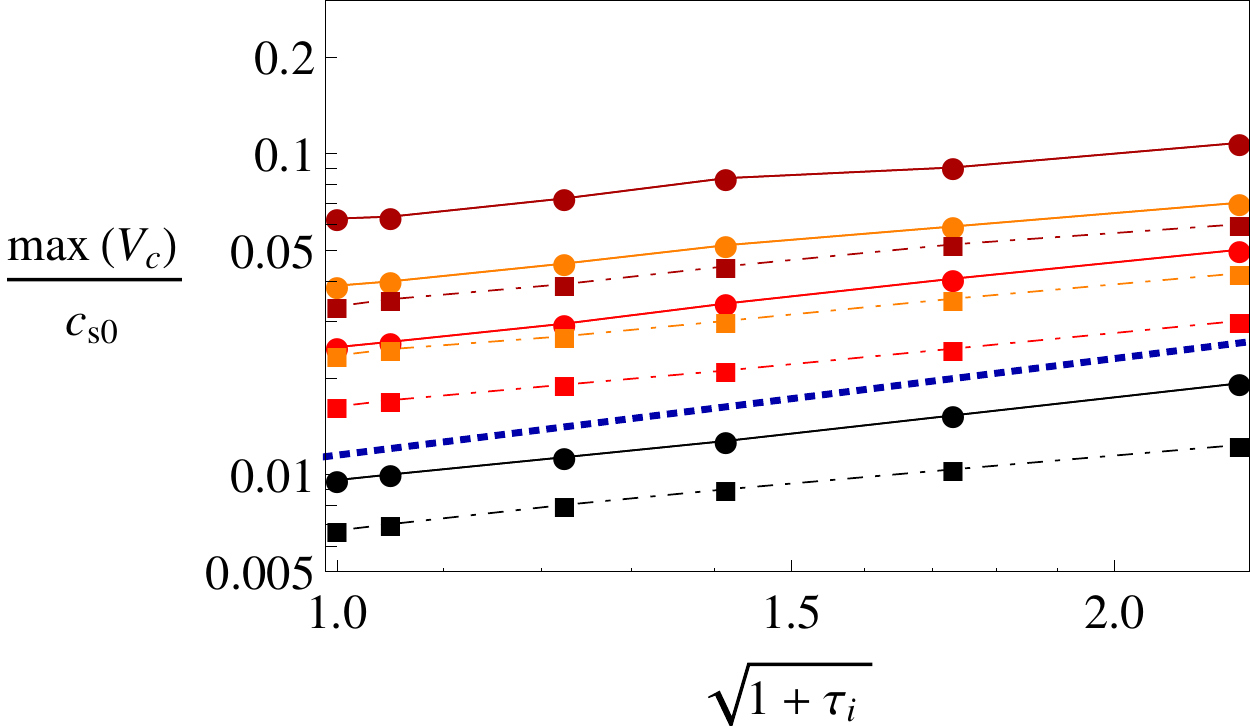} 
\includegraphics[trim = 0px 0px 0px 0px, clip, scale=0.44]{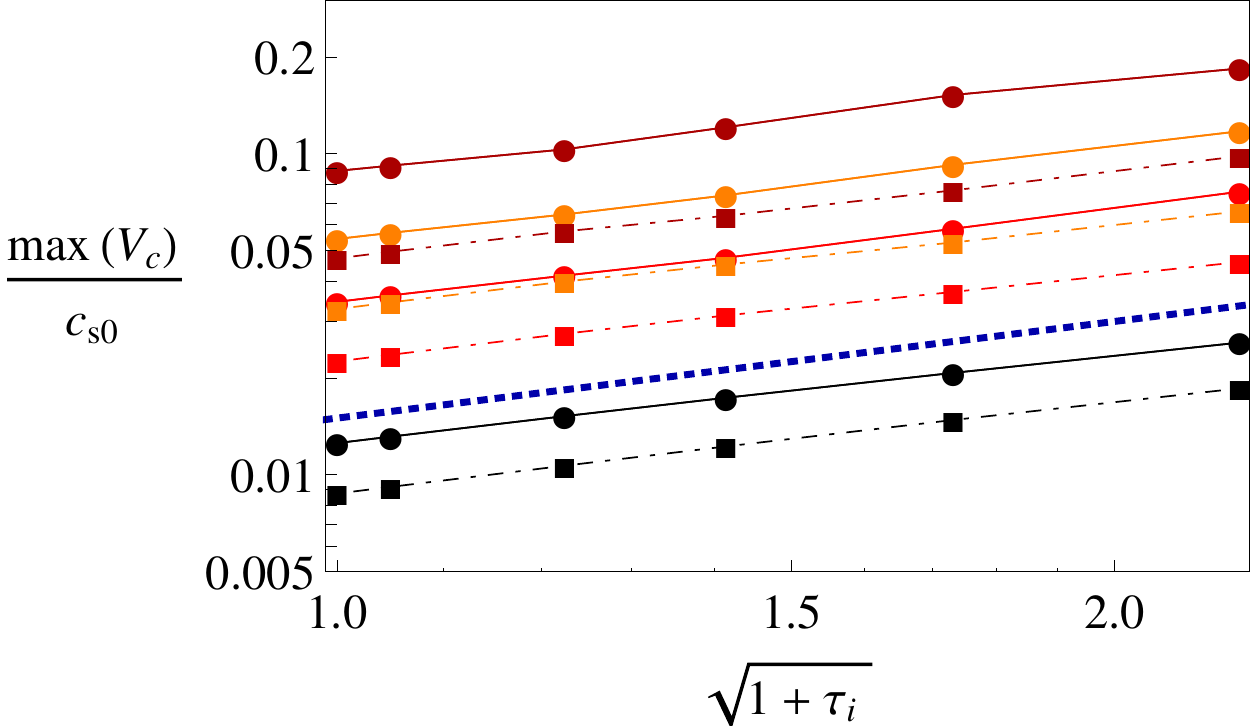} 
\includegraphics[trim = 0px 0px 0px 0px, clip, scale=0.44]{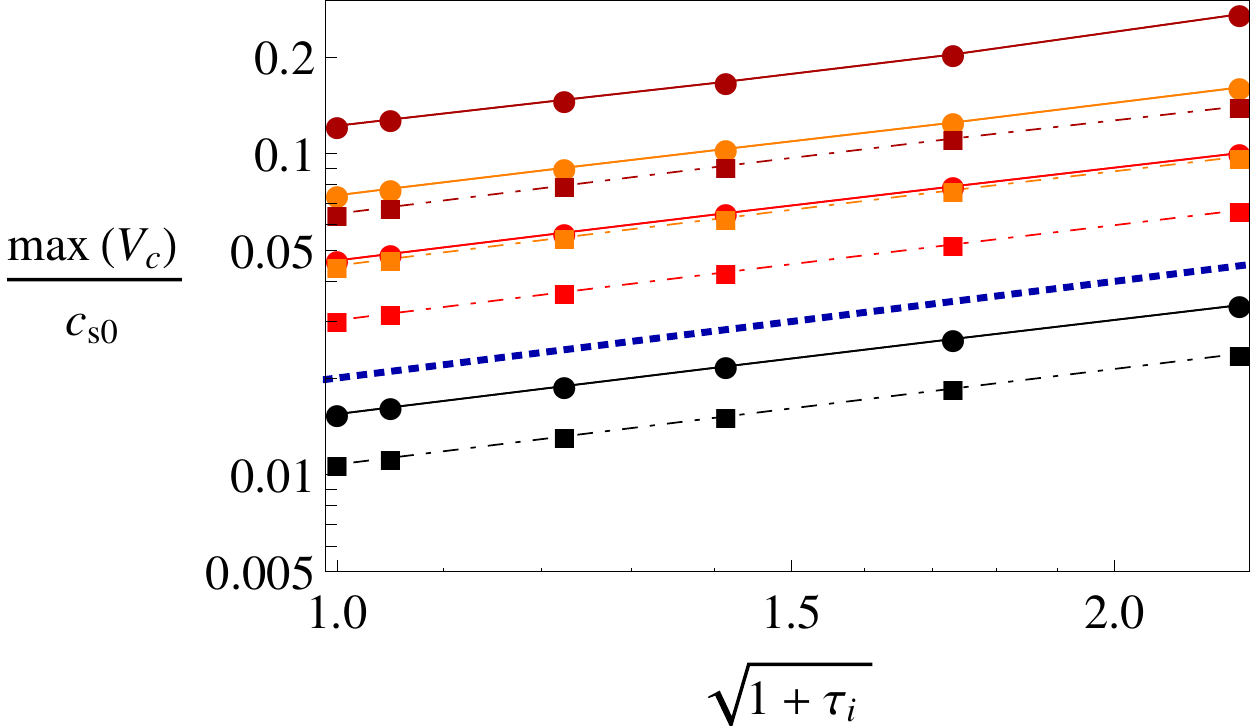}
\caption{The ion to electron background temperature ratio $\tau_i$ dependence of the maximal COM velocity $\textrm{max}(V_{c})$ for 
four different amplitudes $A = \left\{0.1,0.5,1,2\right\}$ (black, red, orange, dark red) and three different cross-field sizes
$\sigma$ are shown in a double logarithmic plot. The solid lines represent thermal blobs and the dot-dashed lines isothermal blobs. The dashed dark blue line represents the $V_\perp/c_{s0}\sim \sqrt{1+\tau_i}$ reference line.}
\label{fig:vCOMmaxvstau}
\end{figure*}
\subsubsection{Hot ions}
\begin{figure}[!ht]
\centering
\includegraphics[trim = 0px 0px 0px 0px, clip, scale=0.62]{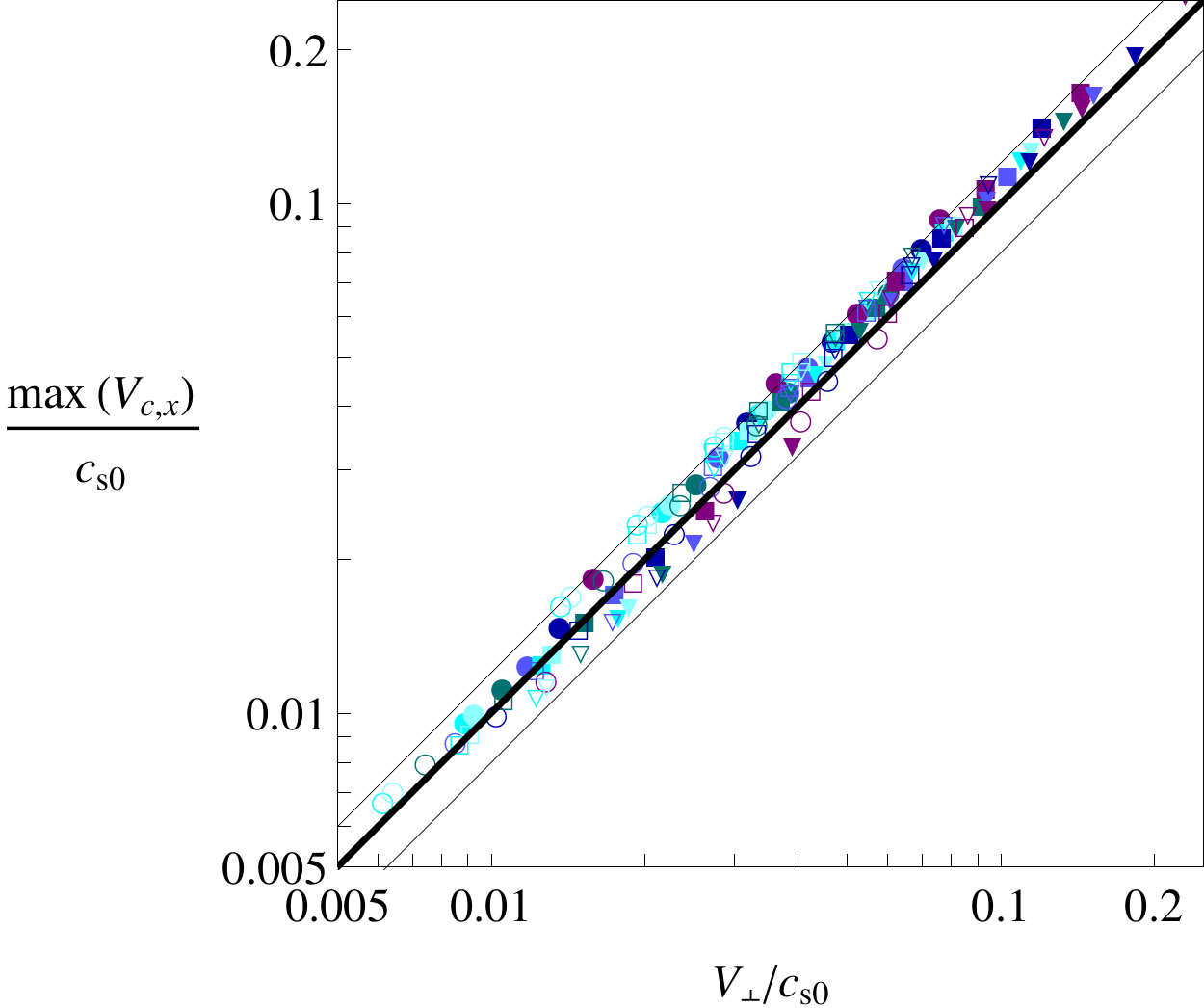}
\caption{A double logarithmic plot of the numerically obtained radial maximal COM velocities $\textrm{max}(V_{c,x})$ versus the velocity scaling estimate $V_\perp$ of~\eqref{eq:velscalingth} is shown. 
The thermal (filled symbols) and isothermal (empty symbols) blob velocities fit excellently to  to the analytical inertial velocity estimate. The grey lines depict a deviation by twenty percent.
The symbols indicate the parameter $\sigma = \left\{5,10,20 \right\} $ (circle, square, down triangle) and the ion to electron background temperature ratio $\tau_i =  \left\{0,0.1,0.5,1,2,4 \right\} $ 
(cyan, light cyan, dark cyan, light blue, dark blue, purple). }
  \label{fig:vnumxvsvtheoxdfthiso}
\end{figure}

\begin{figure}[!ht]
\centering
\includegraphics[trim = 0px 0px 0px 0px, clip, scale=0.60]{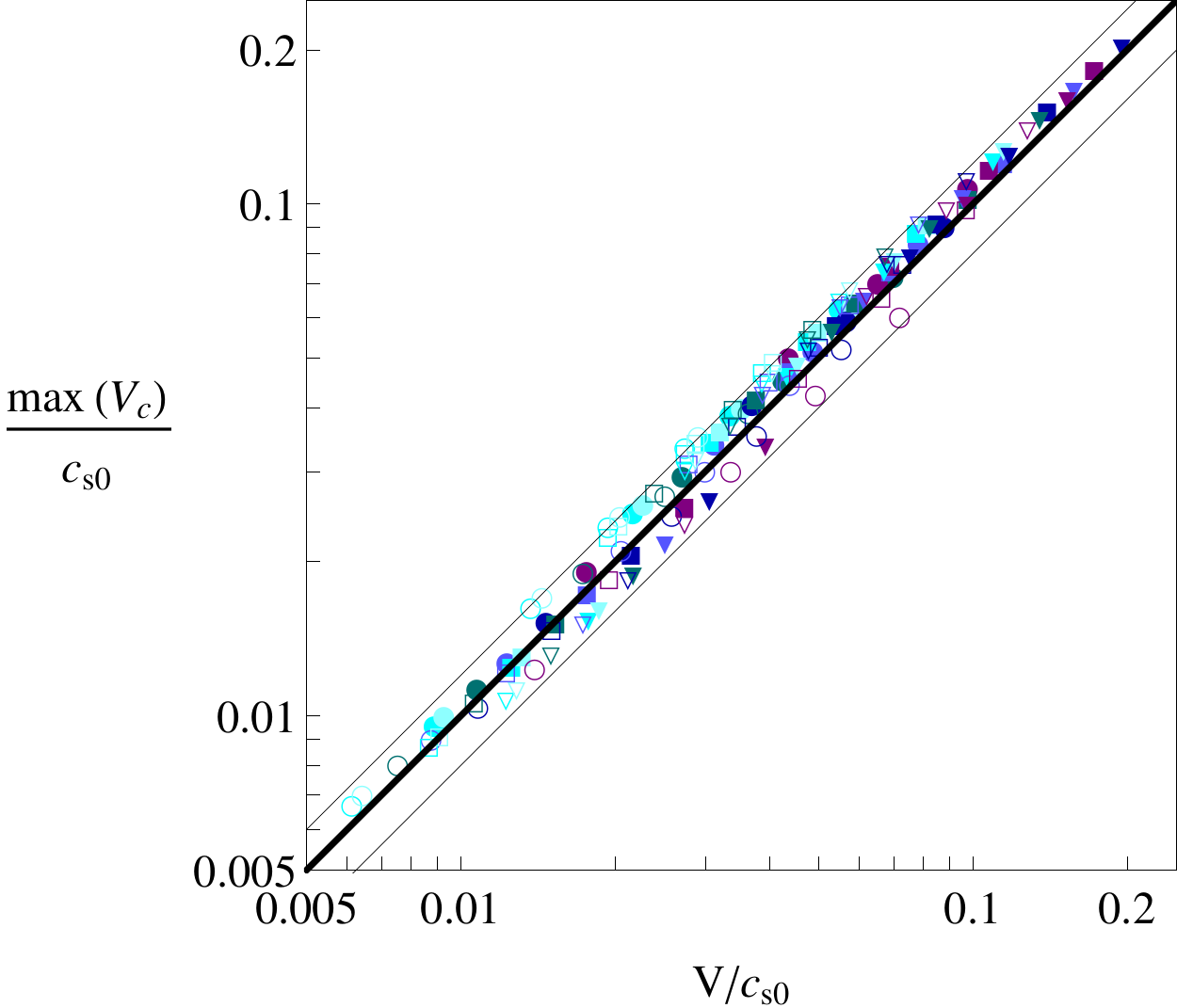}
\caption{A double logarithmic plot of the numerically obtained total maximal COM velocities $\textrm{max}(V_{c})$ versus the empirical velocity scaling estimate $V$ of \ref{eq:totalscaling} is shown. 
The thermal (filled symbols) and isothermal (empty symbols) blob velocities 
fit excellently to the empirical velocity scaling estimate. The grey lines depict a deviation by twenty percent.
The symbols indicate the parameter $\sigma = \left\{5,10,20 \right\} $ (circle, square, down triangle) and the ion to electron background temperature ratio $\tau_i =  \left\{0,0.1,0.5,1,2,4 \right\} $ 
(cyan, light cyan, dark cyan, light blue, dark blue, purple). }
  \label{fig:vtotal}
\end{figure}

\begin{figure}[!ht]
\centering
\includegraphics[trim = 0px 0px 0px 0px, clip, scale=0.65 ]{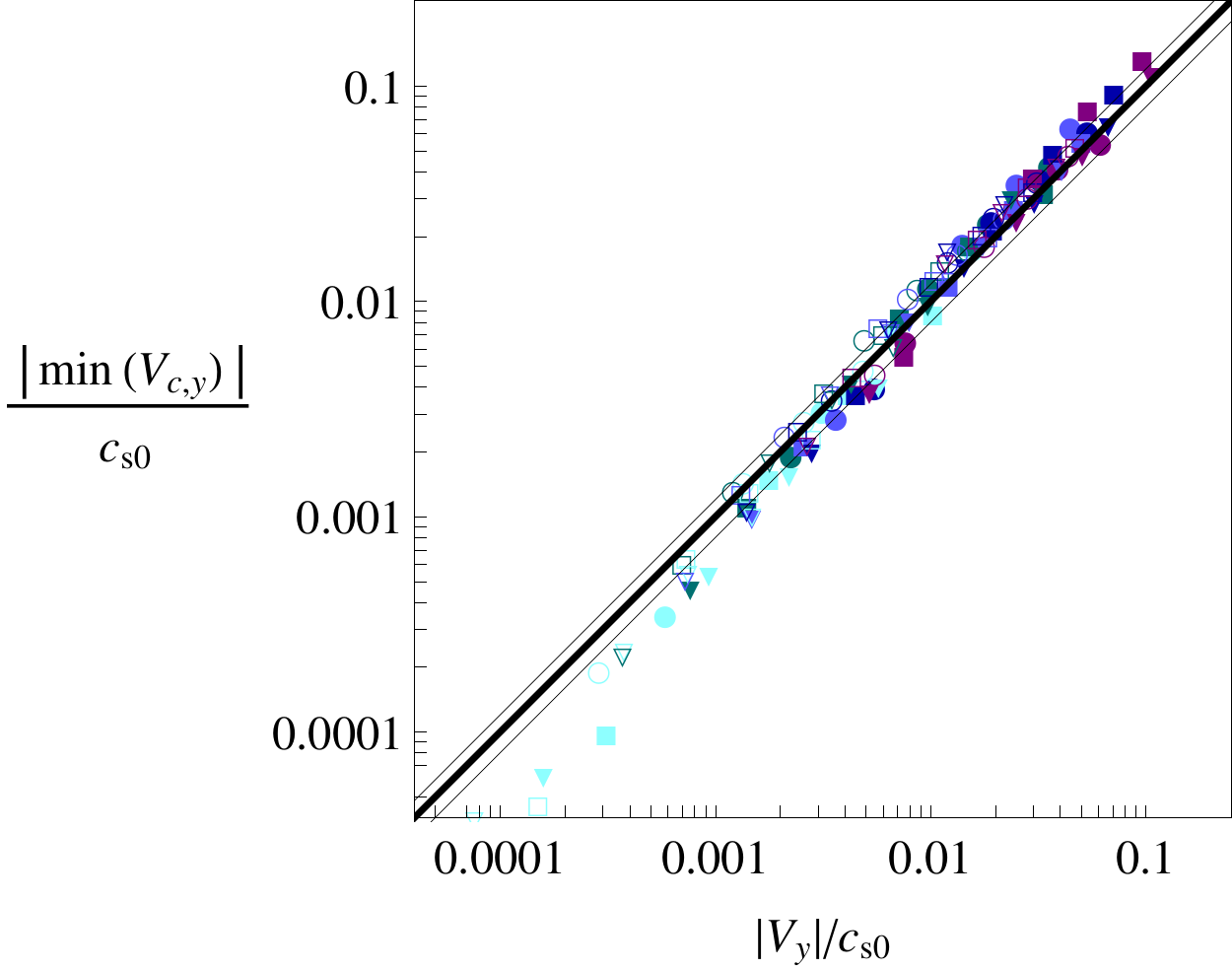}
\caption{A double logarithmic plot of the numerically obtained poloidal minimal COM velocities $\textrm{min}(V_{c,y})$ versus the empirical velocity scaling estimate $V_y$ of \ref{eq:poloidalscaling} is shown. 
The thermal (filled symbols) and isothermal (empty symbols) blob velocities
fit well to empirical velocity scaling estimate. The grey lines depict a deviation by twenty percent.
The symbols indicate the parameter $\sigma = \left\{5,10,20 \right\} $ (circle, square, down triangle) and the ion to electron background temperature ratio $\tau_i =  \left\{0,0.1,0.5,1,2,4 \right\} $ 
(cyan, light cyan, dark cyan, light blue, dark blue, purple). }
  \label{fig:vy}
\end{figure}

Finite ion temperature induces a poloidal COM velocity in the $\vec{\hat{b}} \times\vec{\nabla}B$ direction. 
The poloidal COM velocity $V_{c,y}$ is generated by FLR effects and is further decreased (cf.~\ref{fig:maxtracki1}) if 
dynamic FLR effects arise as a consequence of gradients in ion temperature $t_{i}$ or magnetic field $B$. 
Consequently, in~\ref{fig:velvstimeshots10b} thermal blobs with full FLR dynamics show increased absolute values for the normalised minimal poloidal COM velocities $\textrm{min}(V_{c,y})/V_\perp$ in comparison to blobs with
constant FLR effects especially for high amplitudes.
\\
The behaviour of the normalised radial COM velocities $\textrm{max}(V_{c,x})/V_\perp$ of thermal blobs is similar to blobs with constant FLR effects as depicted in~\ref{fig:velvstimeshots10a}. The 
maximal normalised radial COM velocities $\textrm{max}(V_{c,x})/V_\perp$ are well described by the velocity scaling estimates of~\eqref{eq:velscalingth}. 
As in the cold ion case the  maximal radial COM velocities $\textrm{max}(V_{c,x})$ of thermal blobs 
exceed those of isothermal blobs, which is shown in~\ref{fig:vmaxvstau}.  
However, the ion temperature scaling for the maximal radial and total COM velocity (cf.~\ref{fig:vCOMmaxvstau}) is in line with the theoretical 
estimate $V_{\perp}\sim \sqrt{1+\tau_i}$ of~\eqref{eq:velscalingth} for blobs with high
cross-field size $\sigma=20$. For smaller blob sizes $\sigma=5$ and $\sigma=10$ we observe approximately a decreased scaling $V_{\perp}\sim (1+\tau_i)^{3/8}$ and $V_{\perp}\sim (1+\tau_i)^{5/12} $ respectively.
\\
To verify the inertial velocity scaling law we plot the measured maximal radial COM velocities $\textrm{max}(V_{c,x})$ against the analytical estimates $V_\perp$ of~\eqref{eq:velscalingth}. 
Before we do so we note that all the measured velocities are in the inertial regime. 
The ion pressure dominated RB (iRB) velocity estimate of~\cite{manz13} is not confirmed by our simulations even if some of the parameters fall into this regime. The iRB velocity
underestimates the numerically obtained maximal COM velocities, which is in accordance with~\cite{olsen16}.
In~\ref{fig:vnumxvsvtheoxdfthiso} we prove numerically that the theoretical velocity estimate of the inertial regime (cf.~\eqref{eq:velscalingth}) is recovered over more than an order of magnitude for the thermal and isothermal blobs.
All the derived maximal radial COM velocities lie within 20 percent of the $V_\perp$ estimate, which underlines the usefulness and quality of the inertial scaling law. 
We tested also the global amplitude velocity scaling 
$V_{\perp,g}\sim\sqrt{(\Delta p_{e}+\Delta p_{i})/(n_{e0} + \Delta n_e )}$ of~\cite{kube11,wiesenberger14}. 
However, we obtained  distinct deviations from the numerically obtained maximal COM velocities $\textrm{max}(V_{c,x})$ by the global inertial scaling law $V_{\perp,g}$. 
Hence, we emphasise that the inertial velocity scaling law $V_\perp$ of~\eqref{eq:velscalingth} captures the maximal radial COM velocities $\textrm{max}(V_{c,x})$ best.
\\
In~\ref{sec:prop}, ~\ref{sec:comp} and~\ref{sec:trans} we discussed the parameter $\Theta$ defined in~\eqref{eq:flrstrengthth}, which is a measure of the
influence of FLR effects on blob convection. Specifically we argued that the poloidal and consequently the total transport is related to the FLR strength parameter $\Theta$.
We will now derive an empirical scaling law for the poloidal and total particle transport based on the knowledge of our parameter study. 
The tracks of the COM position $\vec{R}_{c}$ in~\ref{fig:maxtracki1} indicate that for high FLR strength (\eqref{eq:flrstrengthth}) an angle of $-\pi/4$ is taken by the blobs.
Hence, we postulate that this angle is approached by
\begin{align}\label{eq:alpha}
  \alpha \equiv \frac{1}{2} \frac{\tan^{-1}{\left(\Theta+\Theta_s\right)}-\tan^{-1}{\left(\Theta_s\right)}}{1-2\tan^{-1}{\left(\Theta_s\right)}/\pi},
\end{align}
which is a shifted $\tan^{-1}$ function with shift parameter $\Theta_s$. 
The angle $\alpha$ captures the blob compactness $I_c(\alpha) = c_1 \alpha +c_2$ at $t = 4 \gamma_g^{-1}$ up to the constants $c_1$ and $c_2$.
We find that the constants $c_1=\frac{\pi}{2}$, $c_2=\frac{1}{2}$ and the shift parameter $\theta_s=5$ fit the blob compactness $I_C(4 \gamma_g^{-1})$ with a threshold of $4\sigma^2$ (cf.~\eqref{eq:heavi}). 
With~\eqref{eq:alpha} we derive the total and poloidal velocity scaling law
\begin{align}\label{eq:totalscaling}
V \equiv V_\perp/\cos{\left(\alpha\right)}, \\
\label{eq:poloidalscaling}
  V_y \equiv V_\perp\tan{\left(\alpha\right)}.  
\end{align}
In~\ref{fig:vtotal} and~\ref{fig:vy} we compare the derived total and poloidal inertial scaling law $V$ of~\eqref{eq:totalscaling} and $V_y$ of~\eqref{eq:poloidalscaling} with the numerically obtained maximal total COM velocities $\textrm{max}(V_{c})$ and
the numerically obtained minimal poloidal COM velocities $\textrm{min}(V_{c,y})$.
For the total blob velocities we find again agreement up to $\pm 20$ percent. On the other hand, 
the minimal poloidal COM velocities of the thermal and isothermal blobs
are not always within 20 percent of the theoretical estimate of~\eqref{eq:poloidalscaling}. This occurs especially for very small poloidal velocities.
Still,~\ref{fig:vy} underlines the quality of the empirical scaling law.
\section{Discussion and conclusion}\label{sec:conclusion}
Our simulations reveal that variations in the temperature field increase the radial COM blob velocities. 
The maximal radial COM blob velocities obey the inertial velocity scaling law of~\eqref{eq:velscalingth} over the complete parameter scan, 
which encompasses a broad spectrum of typical blob amplitudes $A=\left\{0.1,0.5,1,2\right\}$, ratios of ion to electron background temperatures $\tau_i=\left\{0,0.1,0.5,1,2,4\right\}$
and blob cross-field sizes $\sigma=\left\{5,10,20\right\}$. This coincidence occurs even in the limit of constant temperatures for the complete parameter space.
On the other hand, the total and poloidal particle transport and the compactness of blobs is 
crucially determined by the FLR strength parameter $\Theta$ of~\eqref{eq:flrstrengthth}. 
This parameter describes the transition from the weak to the strong FLR effect regime.
For weak FLR effects ($\Theta \ll 1$) we find by plume-like blob structures in particle density, which primarily travel in the radial direction with maximal radial COM velocities 
captured by the inertial velocity scaling law of~\eqref{eq:velscalingth}.
The time to reach the maximal radial COM velocity is well described by 
the global interchange rate of~\eqref{eq:gammascalingthg}.
For strong FLR effects ($\Theta \gg 1$) compact radially and poloidally moving blob structures are observed in our numerical study. 
The inertial velocity scaling law of~\eqref{eq:velscalingth} estimates the maximal radial COM
velocity over the complete $\Theta$ range.  
In the $\Theta \gg 1$ limit the numerical results suggest that the absolute values of the radial and poloidal COM velocities are equal, leading to the empirical poloidal scaling estimate of~\eqref{eq:poloidalscaling}.
The overall perpendicular particle transport is affected by the inherent particle density of a blob, which in the strong FLR effect regime is
roughly 50 percent higher than in the weak FLR effect regime.
The radial particle transport scales with the square root of the total pressure perturbation (cf.~\eqref{eq:velscalingth}) and thus increases for finite ion temperature.
\\
To gain insight into the transition between various blob regimes and the influence of poloidal particle transport on the divertor heat load 
fully three dimensional computations in X-point geometry are in process of planning. Future plans also involve the study of thermal effects in the sheath connected regime 
and the inclusion  of FLR corrections to the polarisation density. According to~\cite{wiesenberger14} those corrections may slightly enhance the propagation of blobs
with high ratios of ion to electron background temperature, small pressure amplitudes and small blob widths. 
\section{Acknowledgements}
The authors want to thank R. Kube for contributing to this work with fruitful comments.
This work was supported by the Austrian Science Fund (FWF) Y398.
This work has been carried out within the framework of the EUROfusion Consortium and has received funding
from the Euratom research and training programme 2014-2018 under grant agreement No 633053. 
The views and opinions expressed herein do not necessarily reflect those of the European Commission. 
The computational results presented have been achieved in part using the Vienna Scientific Cluster (VSC). 
\appendix
\section{Dissipation of the generalised vorticity density}\label{sec:dissipation}
In the following we will show that the dissipative terms in the generalised vorticity density~\eqref{eq:voreq} reduce approximately to
a hyperdiffusive term of the form $\Lambda_{\mathcal{W}}
\approx  - \nu \nabla_\perp^4 \mathcal{W}$. For the sake of convenience
we show the derivation for common dissipative terms of first order and give the result for second order in the end of this section. \\
The dissipative terms for the gyro-centre densities and temperatures read:
\begin{align}
  \Lambda_{n_e}= \nu \vec{\nabla}_\perp^2 n_e \\
  \Lambda_{N_i}= \nu \vec{\nabla}_\perp^2 N_i \\
  \Lambda_{t_{e}}= \nu \vec{\nabla}_\perp^2 t_{e} \\
  \Lambda_{t_{i}}= \nu \vec{\nabla}_\perp^2 t_{i}
\end{align}
Due to our choice of gyro-centre densities $N$ and temperatures $T$ as dependent variables the ion pressure $p_{i}$ is dissipated according to
\begin{align}
  \Lambda_{p_{i}}&= \nu \left(n_e \vec{\nabla}_\perp^2 t_{i} +  t_{i} \vec{\nabla}_\perp^2 n_e\right)
 \nonumber \\ &
                     = \nu \left(\vec{\nabla}_\perp^2 p_{i} -2 \vec{\nabla}_\perp  t_{i}\cdot\vec{\nabla}_\perp n_e\right).
\end{align}
In the generalised vorticity density~\eqref{eq:voreq} the full expression for the dissipation reads
\begin{align}\label{eq:lambdaomegastar}
  \Lambda_{\mathcal{W}} \equiv\Omega_i\left[   \Lambda_{n_e} -  \Lambda_{N_i} + \vec{\nabla}_\perp^2 \left(\frac{ \Lambda_{p_{i}}}{2 m_i \Omega_i^2} \right)\right].
\end{align}
We rewrite now the first two terms of~\eqref{eq:lambdaomegastar} with the help of~\eqref{eq:iongyrocenterdensity} to
\begin{align}
\Lambda_{n_e} - \Lambda_{N_i} 
      &  \approx        - \nu\vec{\nabla}_\perp^2 \left[ \frac{ \mathcal{W}}{\Omega_i} - \vec{\nabla}_\perp^2  \left(\frac{p_{i}}{2 m_i \Omega_i^2}\right) \right],
      \nonumber \\
\end{align}
and neglect variations in the magnetic field in the last term of~\eqref{eq:lambdaomegastar}
\begin{align}
   \vec{\nabla}_\perp^2\left(\frac{ \Lambda_{p_{i}}}{2 m_i \Omega_i^2} \right) &\approx
  \frac{1}{2 m_i \Omega_i^2}\vec{\nabla}_\perp^2\Lambda_{p_{i}}.
\end{align}
Now it is clear that only for $\Lambda_{p_{i}}= \nu \vec{\nabla}_\perp^2 p_{i} $ the complete dissipation of generalised vorticity density 
reduces to $\Lambda_{\mathcal{W}} =\nu  \vec{\nabla}_\perp^2\mathcal{W} $. This is only possible if the gyro-centre pressure $P$ is evolved instead of the gyro-centre temperature $T$.
With our choice the dissipative term of the generalised vorticity density reduces to:
\begin{align}
\Lambda_{\mathcal{W}}=\nu &\bigg[  \vec{\nabla}_\perp^2 \mathcal{W}
 -\frac{ 1  }{ m_i \Omega_i^2} \vec{\nabla}_\perp^2 \left( \vec{\nabla}_\perp t_{i} \cdot  \vec{\nabla}_\perp n_e\right)\bigg]. 
\end{align}
The derivation with hyperdiffusive terms of second order is analogous and yields 
\begin{align}
\Lambda_{\mathcal{W}}
 =-\nu&  \bigg[   \vec{\nabla}_\perp^4  \mathcal{W}
 -\frac{ 2  }{ m_i \Omega_i^2} \vec{\nabla}_\perp^4 \left( \vec{\nabla}_\perp t_{i} \cdot  \vec{\nabla}_\perp n_e\right)\bigg].
\end{align}
We note here that the artificial dissipative term has no impact on the blob motion in the high Reynolds number regime.
\bibliography{thermalblob}
\end{document}